\documentclass[prl,a4paper,superscriptaddress,twocolumn,showpacs,amsmath,amssymb,floatfix]{revtex4}
\pdfoutput=1

\usepackage{graphicx}
\usepackage{amssymb}
\usepackage{float}
\usepackage{hyperref}

\usepackage{color}

\begin{document}

\title{Nonlinear perturbation of Random Matrix Theory}

\author{Klaus M. Frahm}
\affiliation{\mbox{Laboratoire de Physique Th\'eorique, 
Universit\'e de Toulouse, CNRS, UPS, 31062 Toulouse, France}}
\author{Dima L. Shepelyansky}
\affiliation{\mbox{Laboratoire de Physique Th\'eorique, 
Universit\'e de Toulouse, CNRS, UPS, 31062 Toulouse, France}}

\date{December 22, 2022; Revised: May 17, 2023}

\begin{abstract}
  We consider a system of linear oscillators, or quantum states,
  described by   Random Matrix Theory and analyze how its time evolution
  is affected by a nonlinear perturbation. Our numerical results
  show that above a certain chaos border a weak or moderate nonlinearity
  leads to a dynamical thermalization of a finite number of degrees of freedom
  with energy equipartition over linear eigenmodes as
  expected from the laws of classical statistical
  mechanics. The system temperature is shown to change
  in a broad range from positive to negative values and the dependence of
  system characteristics on the initial injected energy is determined.
  Below the chaos border the dynamics is described by  the Kolmogorov-Arnold-Moser
  integrability. Due to universal features of  Random Matrix Theory
  we argue that the obtained results describe the generic properties of its
  nonlinear perturbation.
  
\end{abstract}

%

\maketitle

In far 1872, 150 years ago,
Boltzmann developed the theory of statistical mechanics
and thermalization originated from the dynamical laws of classical motion
of many-body systems \cite{boltzmann1}.
This result led to the famous Boltzmann-Loschmidt dispute
on a possibility of thermalization and time irreversibility
emerging from the reversible dynamical equations of particle motion
\cite{loschmidt,boltzmann2} (see also \cite{mayer}).
The modern resolution of this dispute
is based on the theory of dynamical chaos
for generic nonlinear systems 
characterized by a positive maximal Lyapunov exponent
and Kolmogorov-Sinai entropy leading to an
exponential instability of motion
(see e.g. \cite{arnold,sinai,chirikov1979,lichtenberg}).
This instability leads to
an exponential growth of errors which breaks time reversibility
(see e.g. an example in \cite{dls1983}). 

The first numerical studies of how ergodicity,  dynamical thermalization
and energy equipartition appear in an oscillator system
with moderate nonlinearity
were reported  by Fermi, Pasta, Ulam in 1955 \cite{fpu1955}.
The conclusion was that ``The results show very little, if any, tendency toward
equipartition of energy between the degrees of freedom.'' \cite{fpu1955}.
It was argued in \cite{zabusky} that in the continuum
limit the Fermi-Pasta-Ulam (FPU) problem
is close to the Korteweg-de Vries equation
with stable soliton solutions
shown to be completely integrable \cite{greene},
as well as the nonlinear Schr\"odinger equation \cite{zakharov}.
In addition, at weak nonlinearity
the FPU $\alpha$-model is close to the completely
integrable Toda lattice \cite{toda,benettin}.
Another explanation of  equipartition absence
in the FPU problem
was given in \cite{chirikovfpu1,chirikovfpu2,livi}
showing that below a certain strength of nonlinear
interactions between oscillator modes the system
is located in the regime of Kolmogorov-Arnold-Moser (KAM)
integrability and only above this border
an overlap of nonlinear resonances takes place with
emergence of chaos and thermalization.
Numerical simulations demonstrated a dynamical thermalization
with energy equipartition reported in \cite{chirikovfpu2,livi}.
Thus, even 50 years after \cite{fpu1955}, various regimes of
nonlinear dynamics of the FPU problem
are actively discussed by the community of dynamical systems \cite{fpu50}
(see e.g. recent \cite{ruffo2022}).
The variety of studies clearly demonstrates that this
model played an important role
in the investigations of nonlinear dynamics
but also that it has multiple specific features
indicating that it does not belong to
a class of generic oscillator systems with nonlinear interactions.

To construct a generic model of many-body oscillator systems
with nonlinear interactions between oscillators we take insight
from quantum mechanics of many-body systems
whose spectral properties are described by Random Matrix Theory (RMT)
invented by Wigner for a description of spectra of complex nuclei, atoms and molecules
\cite{wigner}. At present RMT finds applications in
multiple areas of physics \cite{mehta,guhr}
including systems of quantum chaos whose dynamics is
chaotic in the classical limit \cite{bohigas,haake}.
The properties of RMT eigenvalues and eigenstates were 
established in various studies and are well known.
The RMT eigenstates are ergodic, i.e. uniformly distributed on the 
$N$-dimensional unit sphere, 
and the level spacing statistics is described by the universal RMT distribution
\cite{wigner,mehta,guhr,bohigas,haake}. Due to the linearity of 
the Schr\"odinger equation the time evolution of a wave function 
$\psi$ described by a RMT Hamiltonian 
also describes a time evolution of a system of $N$ linear
oscillators with random linear couplings.
By its own, due to the universal properties
of RMT, it is interesting to understand
how a nonlinear perturbation affects RMT evolution.

With the aim to understand the effects of nonlinear perturbation
of RMT, we consider a simple model described by
the Schr\"odinger equation with a Hamiltonian given by a random
matrix with an additional nonlinear interaction between linear modes:
\begin{equation}
  i\hbar{\partial\psi_n(t) \over\partial t} =  \sum_{n'=1}^N H_{n,n'} \psi_{n'}(t) 
+   \beta \vert\psi_n(t)\vert^2\psi_n(t) \;\;\;\; .
\label{eq1}
\end{equation}
Here $H_{n,n'}$ are elements of an RMT matrix $\hat H$ of size $N$ taken
from the Gaussian Orthogonal Ensemble (GOE) \cite{mehta},
they have zero mean and 
variance $\langle H_{n,n'}^2\rangle = (1+\delta_{n,n'})/(4(N+1))$.
The averaged density of states is given by the 
the semi-circle law 
$dm/dE = \frac{2N}{\pi}\sqrt{1-E^2}$ with typical eigenvalues 
in the interval $E_m\in[-1,1]$ 
(we use dimensionless units with $\hbar=1$),
$\beta$ is a dimensionless constant characterizing 
the nonlinear interaction strength in the original basis $n$.

The eigenmodes of $\hat{H}$ at energies $E_m$ are $\phi_n^{(m)}$ which 
are ergodic with a uniform distribution on the $N$-dimensional 
unit sphere. 
The time evolution of the wave function can be expressed in
the basis of eigenmodes as $\psi_n(t) = \sum^{N}_{m=1} C_m(t)\, \phi_n^{(m)}$
with coefficients $C_m(t)$ giving the occupation probability
$\rho_m =  \langle\vert C_m(t) \vert^2\rangle$ (with 
some long time or ensemble average; see below). The time evolution (\ref{eq1})
has two integrals of motion being the probability norm
$\sum_n \vert \psi_n(t) \vert^2 = 1$ and total energy
$E = \sum_n [<\psi_n(t)|\hat{H}|\psi_n(t)> + (\beta/2) \vert \psi_n(t) \vert^4] $.
At $\beta=0$ the model (\ref{eq1}) can be viewed
as a quantum system or as a classical system of coupled linear
oscillators whose Hamiltonian in the basis of
oscillator eigenmodes is 
${\cal H} = \sum E^{\phantom *}_m C^*_m(t)\, C^{\phantom *}_m(t)$
where $C^{\phantom *}_m, C^*_m$ is a pair of conjugated variables
and $E_m$ plays the role of oscillator frequencies.
Since RMT captures the universal features of quantum and linear oscillator systems
we expect that the model (\ref{eq1}) describes the
universal properties of oscillator systems with chaotic dynamics
induced by weak or moderate nonlinear couplings between oscillators.
We call the model (\ref{eq1}) Nonlinear Random Matrix model (NLIRM).

Above a certain chaos border with $\beta > \beta_c$
a moderate nonlinearity destroys KAM integrability leading to chaotic 
dynamics with a positive maximal Lyapunov exponent $\lambda$.
The nonlinear frequency shift is
$\delta \omega \sim \beta  \vert \psi_n \vert^2 \sim \beta /N$
and, as it was argued in \cite{chirikovyadfiz,dls1993,garcia,mulansky1},
a developed chaos takes place when this shift $\delta \omega$ becomes 
comparable to a typical energy spacing between energies (or frequencies)
of the linear system $\Delta \omega \sim 1/N$. 
Thus  $\delta \omega > \Delta \omega$ implies chaos with the chaos border
 $\beta_c =$ const. $\sim 1$ being independent of system size $N$.

\begin{figure}[t]
\begin{center}
\includegraphics[width=0.42\textwidth]{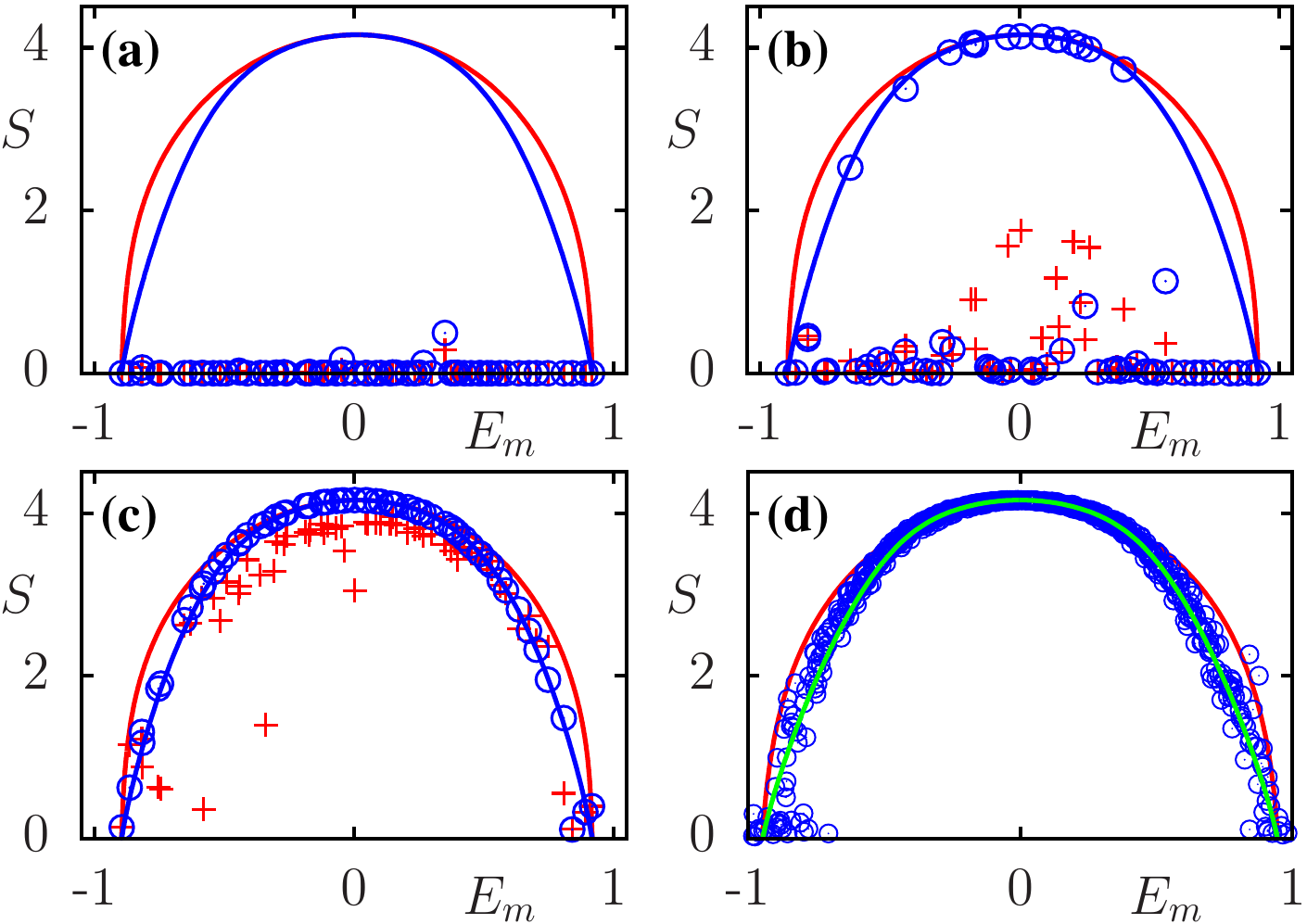}
\end{center}
\vglue -0.3cm
\caption{\label{fig1}
Entropy $S$ versus energy $E_{m}$ of the initial state $m$ at $t=0$
for one RMT realisation at $N=64$ and $\beta=0.02$ (a), $\beta=0.1$ (b) 
and $\beta=1$ (c) or 10 RMT realisations at $\beta=1$ (d). 
The entropy $S$ is computed from  
$\rho_m$  averaged over the time range 
$2^{23}\le t\le 2^{24}$ (blue/black $\circ$) 
and 
$2^{16}\le t\le 2^{17}$ (red/grey $+$ in (a), (b)) or 
$2^{11}\le t\le 2^{12}$ (red/grey $+$ in (c)).
The theory curves  $S(E)$ for BE (red/grey) and EQ (blue/black in (a),(b),(c)
or green in (d)) are from $\rho_m$ values of (\ref{eq2}) 
with $E_m$ values of the used 
RMT realisation (a), (b), (c) or 
a fictitious spectrum according to the semi-circle law in (d)
(where $E_m$ is the solution of 
$m-1/2=M(E_m)$, $m=1,\ldots,N$ with $M(E)$ being the 
integrated density of states). 
}
\end{figure}

The issue of dynamical thermalization
in finite size nonlinear lattices with disorder
was studied in \cite{mulansky1,ermannnjp}.
The time evolution in these systems is
described by the Discrete Anderson 
Nonlinear Schr\"odinger Equation (DANSE)
with hopping between nearby sites.
In the linear case the disorder leads to
Anderson localization of modes \cite{anderson}
which is well visible when the
localization length $\ell$ is smaller than the
system size $N$. In this respect
our RMT model (\ref{eq1}) is rather different
since the linear modes are delocalized and ergodic
in a vector space of dimension $N$.
We expect that our model (\ref{eq1}) is generic and 
captures also certain features of
the models of Bose-Einstein condensate (BEC)
evolution in the chaotic Bunimovich stadium
\cite{stadium} or the Sinai oscillator \cite{sinaioscl}
described by the nonlinear Gross-Pitaevskii equation (GPE) \cite{gpe}.
Indeed, the linear eigenmodes of these systems
have properties of quantum chaos similar to RMT \cite{bohigas,haake}.
There are however also certain differences discussed below.

For the GPE models \cite{stadium,sinaioscl} it is natural to assume that
the dynamical thermalization induced by moderate nonlinearity
leads to the Bose-Einstein (BE) distribution of probabilities $\rho_m$
over quantum levels of the linear system.  In the limit of
high temperature $T$ this distribution is reduced to
a classical energy equipartition (EQ) distribution \cite{mayer,landau}.
For the DANSE type models \cite{mulansky1,ermannnjp}
the quantum Gibbs (QG) distribution was proposed
to explain numerically obtained results.
In fact QG and BE distributions give very close thermalization
properties and we mainly discuss the BE case here.
Thus there are two options for the thermalized distributions
of probabilities $\rho_m$:
\begin{equation}
 \rho_m=\frac{1}{\exp[(E_m-\mu)/T]-1} \; ({\rm BE}), 
\rho_m = \frac{T}{E_m-\mu} \; ({\rm EQ}) .
\label{eq2}
\end{equation}
Here $T$ is the system temperature and
$\mu(T)$ is the chemical potential dependent on temperature.
The parameters $T$ and $\mu$ are determined
by the norm and energy conservation
$\sum_m \rho_m =1$ 
and  $\sum_m E_m \rho_m =E$
(for $E$ we assume the case of weak or moderate
nonlinearity which gives only a weak
contribution to the total energy).
The entropy $S$ of the system is determined by
the usual relation \cite{mayer,landau}:
$S= - \sum_m \rho_m \ln \rho_m$
with the implicit  theoretical dependencies on temperature
$E(T)$, $S(T)$, $\mu(T)$.
The derivation of (\ref{eq2}) is given in 
Supplementary Material (SupMat).

Based on classical statistical mechanics  \cite{mayer,landau}
the dynamical thermalization should lead to the EQ distribution
(\ref{eq2}) since DANSE, GPE  \cite{mulansky1,ermannnjp,stadium,sinaioscl}
and NLIRM (\ref{eq1}) models describe classical nonlinear fields
without second quantization. In contrast, in
\cite{mulansky1,ermannnjp,stadium,sinaioscl} it was argued
that a moderate nonlinearity plays a role of an effective
nonlinear thermostate that leads to quantum BE or QG
distributions (\ref{eq2}).

Of course, both BE and EQ approaches (\ref{eq2})
give different thermal characteristics
leading to a contradiction discussed in
detail in \cite{ermannnjp,stadium,sinaioscl}.
The main argument in favor of the BE (or the QG) ansatz
was based on a reasonably good agreement
of numerical data for entropy vs energy
with the theoretical thermal dependence $S(E)$ given
by the BE (or QG) ansatz. The quantities $S$ and $E$ are
extensive (self averaging) and it
was argued that their analysis is more preferable as 
compared to the direct study of the 
strongly fluctuating probabilities
$\rho_m$ \cite{mulansky1,ermannnjp,stadium,sinaioscl}.
Here we show that the ergodicity of RMT eigenstates 
of $ \hat{H}$ allows to reduce significantly the 
fluctuations and to obtain stable results
for $\rho_m$ that are clearly described by the EQ ansatz (\ref{eq2}).

The numerical integration of (\ref{eq1})
is done with the symplectic scheme of order 4 
\cite{forest,integrator1,integrator2}
using a step size $\Delta t=0.1$ 
up to maximal times $t=4 \times 10^6$ - $1.3\times 10^8$
with exact norm conservation, energy conservation
with accuracy $\sim 10^{-8}$ and for the GOE matrix size $N=64$ 
(see SupMat for more details and results for 
other values $N=32,128,256,512$). 
As initial condition, we choose an eigenmode $\phi_n^{(m)}$ of 
$\hat H$ at some index $m$ (sometimes also noted $m_0$) such 
that the energy remains close to the initial energy $E\approx E_m$. 
Examples of the time dependence $S(t)$ are shown in SupMat Fig.~S1
demonstrating a steady-state regime reached at times $t > 10^4$ 
for $\beta=1$. 
The obtained dependence $S(E)$ is shown in Fig.~\ref{fig1} at different 
$\beta$ values for a specific RMT realisation and two time scales 
and also for 10 RMT realisations at $\beta=1$. 
At small values $\beta = 0.02, 0.1$ the system is close to
an integrable KAM regime \cite{chirikov1979,lichtenberg}
while at $\beta=1$ essentially all modes are
thermalized (see Fig.~\ref{fig1}, SupMat Fig.~S2  and
additional material in \cite{ourwebpage}).
These results show that the critical border for thermalization is 
located at $\beta_c \sim 0.1$
independent of $N$. However, the exact determination of $\beta_c$
is a rather complicated task due to the presence of
many-body nonlinear effects like
e.g. the Arnold diffusion \cite{chirikov1979,lichtenberg}.
Also at the spectral borders $E \approx \pm 1$
the spacing between energies $E_m$ increases according to
the semicircle law \cite{mehta} and therfore 
it is more difficult to reach thermalization there.

\begin{figure}[t]
\begin{center}
\includegraphics[width=0.42\textwidth]{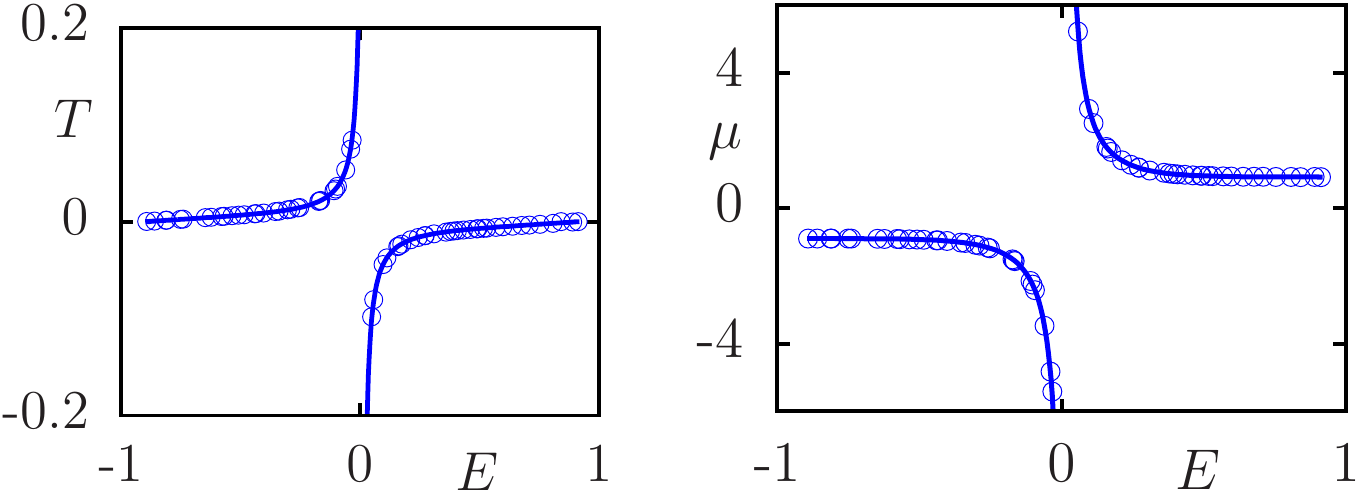}
\end{center}
\vglue -0.3cm
\caption{\label{fig2}
  Dependence of $T$ and $\mu$ on energy $E$
  for EQ ansatz (\ref{eq2}) (curves);
  data points are for $\beta=1$, $N=64$
  and time range $2^{23}\le t\le 2^{24}$,
  with  $T$ and $\mu$ determined from  norm  and 
  numerical entropy values
  (same RMT realisation as in Fig.~\ref{fig1} (c)).
}
\end{figure}

An important feature of Fig.~\ref{fig1}
is that the theory curves $S(E)$ obtained
with the BE and the EQ ansatz (\ref{eq2}) are rather
close to each other. Thus due to fluctuations
of numerical data for $S(E)$ it is difficult
to determine which theory BE or EQ describes better the numerical data. 
However, the data points are significantly closer to the BE-curve, especially 
for moderate energies $|E|\approx 0.5-0.8$ where both curves are somewhat 
different
(the difference between the QG and BE $S(E)$ curves, 
not visible on graphical precision, is $\sim 0.003$ at 
the spectral borders and much smaller at other $E$ values,
so that we discuss mainly the BE case).

For the EQ ansatz the dependencies $T(E)$, $\mu(E)$, obtained 
by the solution of the equations for energy 
and norm for a given RMT spectrum, 
are shown in Fig.~\ref{fig2} (SupMat Fig.~S3 for the BE ansatz) 
for the thermalized regime at $\beta=1$. The numerical points obtained
from $E$ and norm values are by definition exactly located on the 
theory curves.
If instead of $E$ we use the numerical data of $S$ 
then the points slightly deviate from the theory (Fig.~\ref{fig2} and
SupMat Fig.~S3)
but  $T$ and $\mu$ values 
themselves are drastically different between BE and EQ cases.

\begin{figure}[t]
\begin{center}
\includegraphics[width=0.42\textwidth]{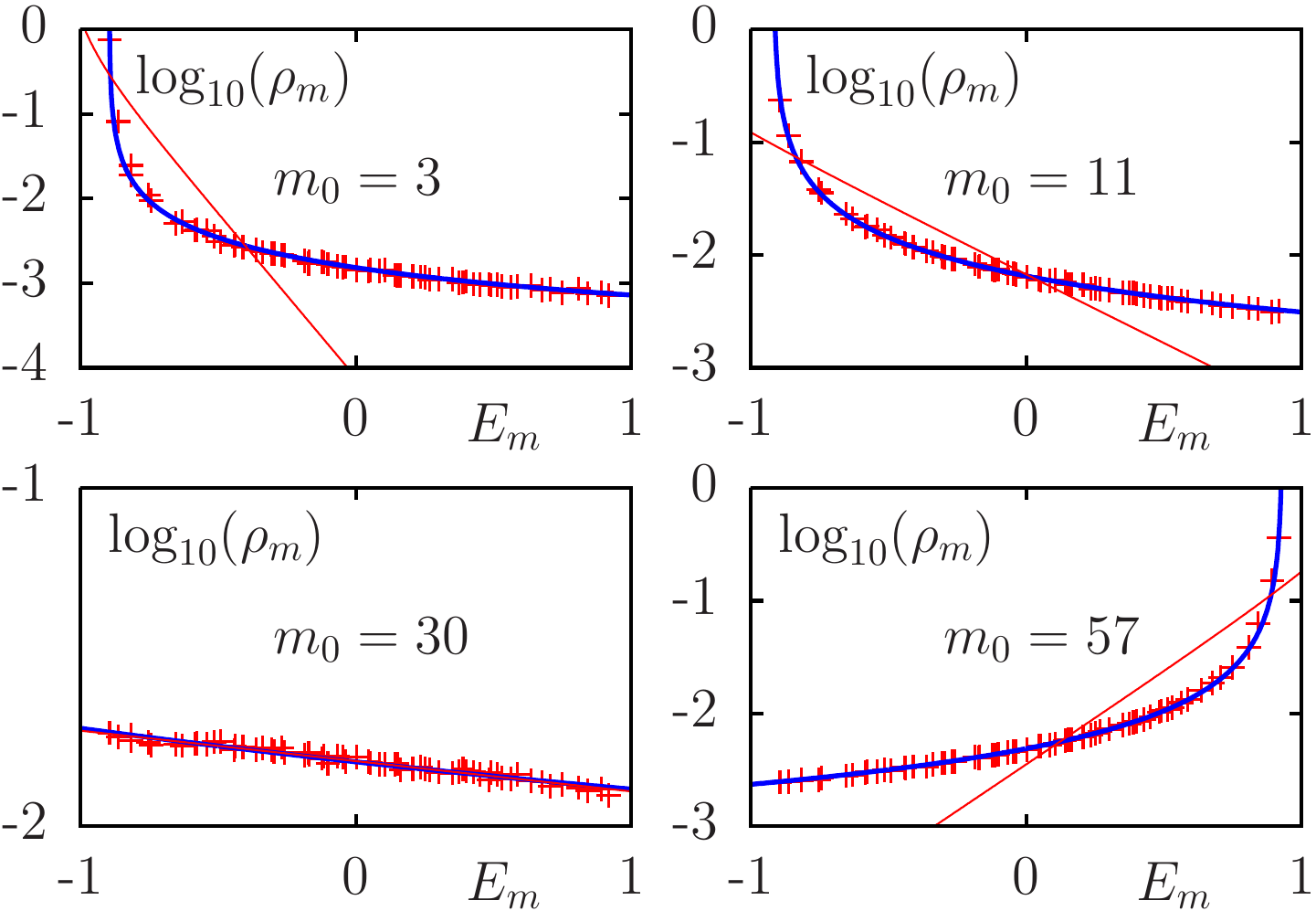}
\end{center}
\vglue -0.3cm
\caption{\label{fig3}
Dependence of $\rho_m(E_m)$ on $E_m$ for 4 initial 
states at $m_0=3, 11, 30$ and $57$ with negative temperature $T<0$;
here $\beta=1$, $N=64$ and time average range $2^{23}\le t\le 2^{24}$. 
The blue curve shows theory of  
 EQ ansatz with  
 with $\rho_{\rm EQ}(E)=T/(E-\mu)$.
 The red line shows BE ansatz theory
 $\rho_{\rm BE}(E)=1/(\exp[(E-\mu)/T]-1)$;
 $T, \mu$ theory (\ref{eq2}) 
 values are given in SupMat Fig.~S4 for BE and EQ cases.
}
\end{figure}

\begin{figure}[t]
\begin{center}
\includegraphics[width=0.42\textwidth]{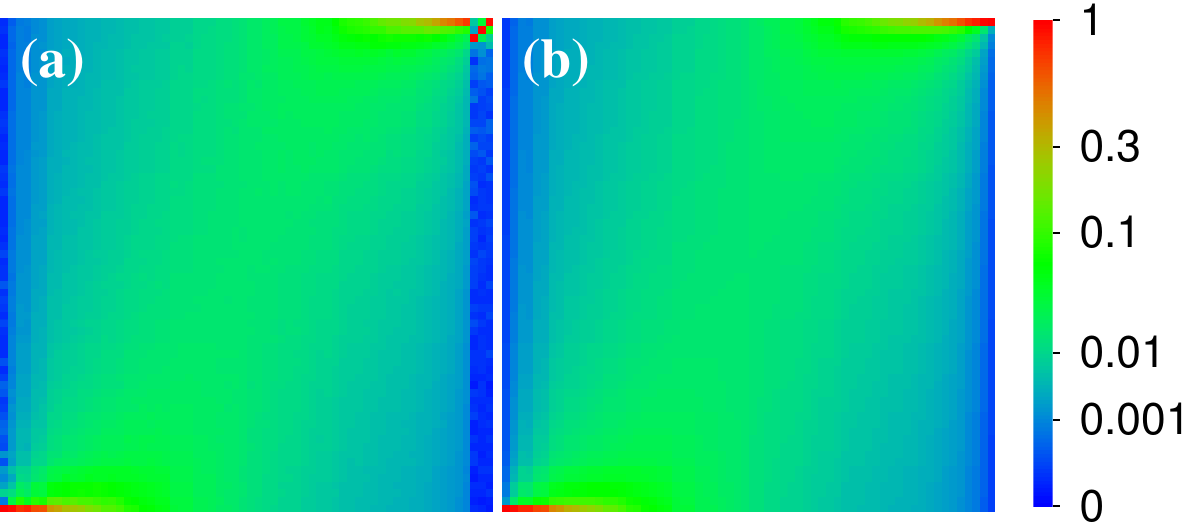}
\end{center}
\vglue -0.3cm
\caption{\label{fig4}
Density plot of $\rho_m$ for parameters of Fig.~\ref{fig3}  
with initial state index $1\leq m_0 \leq 64$ in the $x$-axis and
$1 \leq m \leq 64$ in the $y$-axis. The colorbar 
shows $\rho_m$ values in a nonlinear scale 
to increase the visibility of small $\rho_m$ values.
Panel (a) shows numerical data for $\beta=1$, $N=64$; 
panel (b) shows the EQ ansatz 
$\rho_{\rm EQ}(E_m)$ (see also Fig.~\ref{fig3}).
}
\end{figure}

The most direct way to distinguish between BE and EQ cases
is to compare the probability dependence $\rho_m(E)$
with the theory (\ref{eq2}).
Such a comparison is shown in Fig.~\ref{fig3} for 4 initial states at 
m=$m_0$, $\beta=1$ and $N=64$
(more data are  in SupMat Fig.~S4  and  \cite{ourwebpage}).
The dynamical thermalization clearly follows the EQ ansatz and not at all 
the BE one, except for an initial state at $E_{m_0}\approx 0$ where 
both approaches are equivalent. This observation is in agreement 
with the classical statistical mechanics \cite{mayer,landau}. 
The probabilities $\rho_m$ for all initial
energies $E_{m_0}$ are shown in Fig.~\ref{fig4}
with a good agreement between numerical data and the 
EQ ansatz (see \cite{ourwebpage} for figures as Fig.~\ref{fig3} 
for all $m_0$ values).
The statistical distribution $p(x)$ of fluctuations of the rescaled 
quantity $x=(E_{m_0}-\mu)|C_m(t)|^2/T$ (with $\mu$, $T$
from the EQ ansatz for the energy $E_{m_0}$) 
also follows the  Boltzmann law $p(x)=\exp(-x)$
(see SupMat Fig.~S5). 

\begin{figure}[t]
\begin{center}
\includegraphics[width=0.42\textwidth]{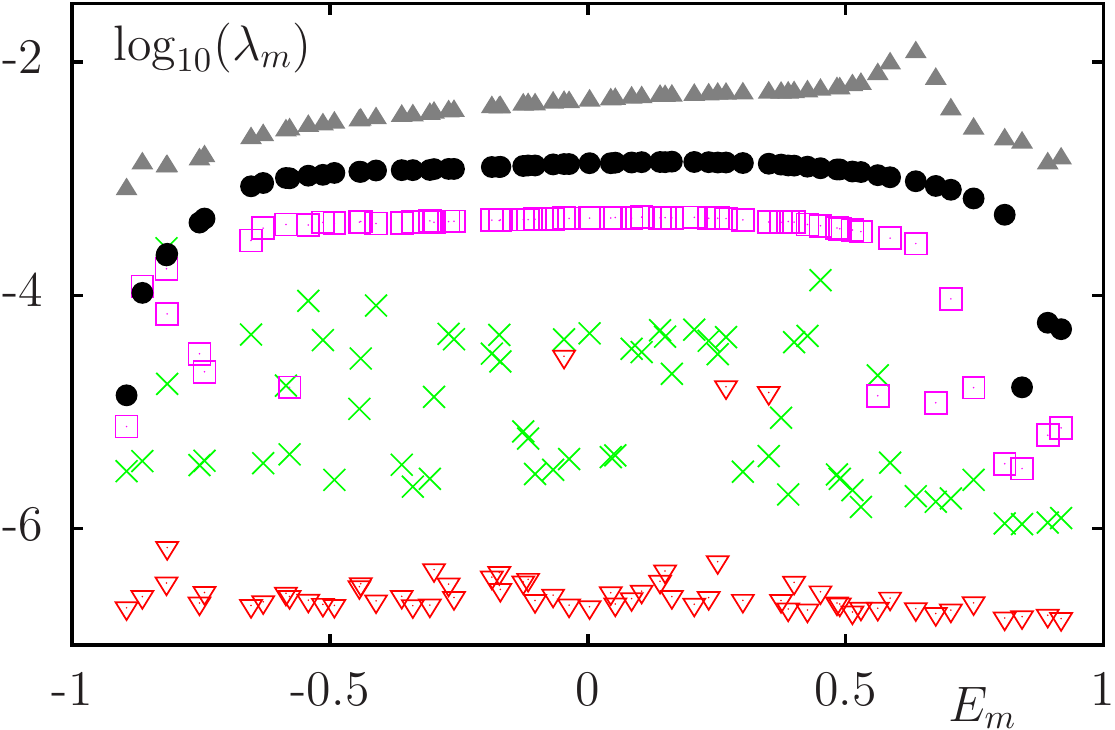}
\end{center}
\vglue -0.3cm
\caption{\label{fig5}
Lyapunov exponent $\lambda_{m}$ dependence on $E_{m}$ with 
$m$ being the index of the initial state for $N=64$; $\lambda_{m}$ 
is determined from the fit $\ln\|\Delta\psi(t)\|
=a+b\ln(t)+\lambda_{m}\,t$ 
for $\beta=2$ (grey $\blacktriangle$; top),
$1$ (black $\bullet$),
$0.5$ (pink $\square$),
$0.1$ (green $\times$) at $t\le 2^{22}$;
$\beta=0.02$ for $t\le 2^{27}$ (red $\triangledown$; bottom).
}
\end{figure}

In Fig.~\ref{fig5} we show the energy dependence of the 
maximal positive Lyapunov exponent $\lambda_m$ on energy $E_m$ 
of initial state  $m$ for different $\beta$ values
(more data are  in SupMat Figs.~S6-S10 and \cite{ourwebpage}).
In the thermalized phase $\beta=1$ we have a smooth variation of $\lambda$
with $E_m$ while below or close to the thermalization border
at $\beta =0.1$ high $\lambda_m$ values  appear only at specific $E_m$ 
values. We attribute this to the existence of triplets of energies
with very close $E_m$ values. Indeed, in a hypothetic case of
3 equal $E_m$ values the KAM theory is not valid and
developed chaos exists at arbitrary small $\beta$ values
as it shown in \cite{chirikovyadfiz,mulansky2}.
Nonetheless, in RMT there is level repulsion and
double or triple degeneracies are 
forbidden leaving place only to quasi-degeneracy of
levels so that KAM becomes valid at $\beta \rightarrow 0$.
Thus for  $\beta=0.02$ we have typically 
$\lambda_m$ approaching to zero with increasing time.
Our preliminary results show that in the thermal phase
at larger $|T|$ (if $E_m\approx 0$) we have an approximate dependence
$\lambda \sim \beta^\eta/N^\nu$
with $\eta \approx 1.52, \nu \approx 1.89$
(see SupMat Figs.~S6-S10). However,
the Lyapunov exponent dependence
on $\beta$ and $N$ requires further more detailed studies. 

Finally, we discuss the reasons
why the nature of thermal equipartition, BE or EQ, was so difficult
to establish in previous studies
\cite{mulansky1,ermannnjp,stadium,sinaioscl}.
One of them is the proximity of $S(E)$ curves for both approaches. 
At the same time the direct
determination of the $\rho_m(E)$ dependence
is rather difficult due to significant
fluctuations, as it was pointed out previously.
These fluctuations are especially large
for the DANSE case at a large disorder
($W=4$ in \cite{mulansky1}) when the localization length $\ell$
is significantly smaller than system size $N$
($\ell/N \approx 0.1$ at $N=64$).
We illustrate this in SupMat Figs.~S11-S12
showing that at smaller disorder $W=2$
with larger localization length $\ell$ the fluctuations of $\rho_m$
are reduced and at long times we have
an agreement of $\rho_m(E)$ with the EQ ansatz 
and strong deviations from the BE ansatz.
For NLIRM model (\ref{eq1}) the linear eigenmodes
are ergodic, i.e. no localization, and the fluctuations of  $\rho_m(E)$ 
are significantly reduced that allows to  distinguish clearly
between EQ and BE cases. 

The cases of GPE in the Bunimivich stadium \cite{stadium}
and the Sinai-oscillator trap \cite{sinaioscl}
are somewhat different. Indeed, in these models
the spectrum of the linear system is unbounded
so that, even if linear eigenstates are in the
quantum chaos regime, the probability spreading to
high energies is rather slow due to small
coupling transitions induced by nonlinearity
between states with significantly different 
energies. Thus in these systems
there is a formation of a relatively
compact probability packet at low energies
which spreads to high energies very slowly
in time. Such an energy  
packet of  $\rho_m$ gives $S(E)$
values compatible with the curve of the BE ansatz
however the fluctuations of $\rho_m(E)$
are very strong with a significant difference
from the BE distribution at high energies
(see e.g. Fig.5 in \cite{stadium} and
Figs.~8,11 in \cite{sinaioscl}).
To analyze these features in more detail, 
we add to the diagonal RMT matrix 
element $H_{n,n}$ an additional diagonal energy 
$fn$ with a constant $f>0$. Then the  variation
of linear energies  $fN$ becomes rather large 
and exceeds significantly those of the RMT case. 
The results for this model at $\beta=1$, $f=0.25$ 
show that at times $t = 2^{15}$ for $N=32$ 
(or $t=2^{20}$ for $N=64$) 
the probabilities $\rho_m(E)$ form a compact
packet of approximate BE shape and the EQ thermal distribution
is reached (with fluctuations) only at very large
times $t=2^{27}$ (see SupMat Figs.~S13,~S14).
Such large time scales 
were out of reach in \cite{stadium,sinaioscl}
due to the complexity of the numerical integration of GPE.

In conclusion, we showed that a nonlinear perturbation
of RMT leads to dynamical thermalization
with energy equipartition corresponding to
the laws of classical statistical mechanics \cite{mayer,landau}.
Such a thermalization appears due to dynamical chaos
in finite systems
with moderate or large number of degrees of freedom
at weak or moderate perturbation of a linear RMT system.
At very weak perturbations the system dynamics
is characterized by a quasi-integrable KAM regime.
We argue that the proposed NLIRM model
captures the generic features
of dynamical thermalization in
systems weakly perturbed by classical nonlinear fields 
and does not depend on the specific form of the nonlinear term 
(see detailed discussion in SupMat and Figs.~S15, S16 there).
Of course, for finite many-body quantum systems with second
quantization the interactions lead
to quantum dynamical thermalization
and distributions of Bose-Einstein for bosons
or Fermi-Dirac for fermions, as it has been
demonstrated in numerical studies \cite{bosons}
and \cite{fermions1,fermions2,fermions3} respectively.

NOTE ADDED: After submission of this work
a  dynamical thermalization at
negative temperature in EQ (2) 
has been observed in optical fibers \cite{fiber},
see also discussion in SupMat.

\noindent {\bf Acknowledgments:}
This work has been partially supported through the grant
NANOX $N^o$ ANR-17-EURE-0009 in the framework of 
the Programme Investissements d'Avenir (project MTDINA).
This work was granted access to the HPC resources of 
CALMIP (Toulouse) under the allocation 2022-P0110.


\newpage

\textheight=21cm


\setcounter{figure}{0} \renewcommand{\thefigure}{S\arabic{figure}} 
\setcounter{equation}{0} \renewcommand{\theequation}{S.\arabic{equation}} 
\setcounter{page}{1}

\noindent{{\bf Supplementary Material for\\
\vskip 0.2cm
\noindent{Nonlinear perturbation of \\
Random Matrix Theory}\\}
\bigskip

\noindent by
K.~M.~Frahm and D.~L.~Shepelyansky\\
\noindent Laboratoire de Physique Th\'eorique, 
Universit\'e de Toulouse, CNRS, UPS, 31062 Toulouse, France
\bigskip

See also [39], corresponding to 
\url{https://www.quantware.ups-tlse.fr/QWLIB/nonlinrmt/}, 
(Accessed Dec 22, 2022), 
for additional and more detailed figures.

\newcommand{\formula}[1]{
        \begin{align*}
                #1
        \end{align*}
}

\newcommand{\lformula}[1]{
        \begin{align}
                #1
        \end{align}
}
\def\C{{\mathbb C}}
\def\K{{\mathbb K}}
\def\R{{\mathbb R}}
\def\Q{{\mathbb Q}}
\def\Z{{\mathbb Z}}
\def\N{{\mathbb N}}
\def\ra{\Rightarrow\quad} 
\def\folgt{\quad\Rightarrow\quad} 

\section{Statistical classical theory}

The nonlinear Schr\"odinger equation (1) has two 
integrals of motion. By neglecting the energy of the weak
nonlinear term $\sim \beta/N$
and assuming global chaos and ergodicity, we expect that the 
system becomes ergodic or thermalizes on the manifold fixed by 
the two constraints~:
\formula{\sum_m E_m|C_m|^2=E\quad,\quad \sum_m |C_m|^2=1}
where $C_m$ are the coefficients of the state in the expansion 
of the eigenbasis of the matrix $\hat H$. 
This situation corresponds in principle to a micro canonical ensemble 
with an additional constraint which is technically quite complicated.
One can use $|C_1|^2=1-\sum_{m=2}^N |C_m|^2$ to remove the first 
coordinate $C_1$ from the phase space to obtain a pure micro canonical 
ensemble for $C_2,\ldots,C_N$ with:
\formula{E-E_1=\sum_{m=2}^N (E_m-E_1)|C_m|^2}
but there is still the condition $\sum_{m=2}^N |C_m|^2=1-|C_1|^2\le 1$ 
which creates technical complications. For small temperature or energy 
(with $E$ being close to $E_1$, assuming an ordered 
eigenvalue spectrum $E_1<E_2<\ldots< E_N$) one can neglect this condition 
and in this case it is not difficult to show by standard text book 
techniques of statistical physics that in the limit $N\gg 1$ 
the marginal distribution of a field $C_m$ (integrating out the 
other fields of the micro-canonical ensemble) is a (complex) Gaussian 
\formula{p(C_m)\sim \exp\left(-\frac{(E_m-E_1)|C_m|^2}{T_{mc}}\right)}
with the micro-canonical temperature $T_{mc}=(E-E_1)/N$ 
and providing the equipartition average~:
$\rho_{m,mc}=\langle |C_m|^2\rangle =T_{mc}/(E_m-E_1)$. 

However, for larger energies the additional inequality 
for the coefficients $C_m$ cannot be neglected. Therefore, we 
treat the system as a grand-canonical ensemble, which is equivalent 
for $N\gg 1$. 
In this approach the fields $C_m$ can freely fluctuate and the 
constraints are only verified in average. The classical 
grand canonical partition function is given by
\formula{Z&=\int \prod_m d^{\,2} C_m \,\exp
\left(-\frac{1}{T}\sum_m (E_m-\mu )|C_m|^2\right)\\
&\sim T^N\prod_m\frac{1}{E_m-\mu}\folgt\\
\ln(Z)&=N\ln(T)-\sum_m\ln(E_m-\mu)+\text{const.}
} 
with two parameters being the (grand canonical) temperature $T$ 
and the chemical potential $\mu$ which are determined by 
the implicit equations 
\lformula{\label{eq_imp}1=\sum_m \rho_m\quad,\quad E=\sum_m E_m\rho_m}
with $\rho_m$ being the statistical average~:
\formula{\rho_m=\langle|C_m|^2\rangle=\frac{T}{E_m-\mu}\equiv 
\rho_{EQ}(E_m)\ .}
Here we have either $T>0$ and $\mu<E_1$ or $T<0$ and 
$\mu>E_N$ in order to have well defined Gaussian integrals 
in the partition function and only solutions for $T$ and $\mu$ 
satisfying this condition are valid. From 
\lformula{\nonumber
E-\mu&=\left\langle\sum_m (E_m-\mu)|C_m|^2\right\rangle
=T^2\frac{\partial\ln(Z)}{\partial T}\\
\label{eq_T}
&=T^2\frac{N}{T}\folgt T=\frac{E-\mu}{N}
}
we find that $\mu$ is a solution of the implicit equation:
\lformula{\label{eq_mu}1=T\sum_m\frac{1}{E_m-\mu}=
\frac1N\sum_m\frac{E-\mu}{E_m-\mu}\ .}
For a given value of $E$ and a given spectrum $E_m$ this equation 
can be solved numerically by standard techniques and using (\ref{eq_T}) 
we also obtain $T$ once $\mu$ is known. 
Depending on the sign of $E-\sum_m E_m<0$ (or $>0$) we have either 
$\mu<E_1$ and $T>0$ (or $\mu>E_N$ and $T<0$) as unique and physically 
valid solution (mathematically there are typically many other but invalid 
solutions of (\ref{eq_mu}) in the interval $E_1<\mu<E_N$).
Once $\mu(E)$ and $T(E)$ are known one can 
use $\rho_m$ to compute the entropy 
\formula{S_{EQ}(E)&=-\sum_m \rho_{EQ}(E_m)\ln(\rho_{EQ}(E_m))\ .}
This expression was used to compute the theoretical $S(E)$ curves 
in the equi-partition 
approach based on the grand-canonical classical theory shown 
in Figs.~1, \ref{fig_SE2}, \ref{fig_DANSE_SE}, 
\ref{fig_DIAG_SE} for various examples. 

We mention that the grand canonical temperature (\ref{eq_T}) 
is similar to the micro-canonical temperature 
if we replace $E_1\to\mu$ and it is not difficult to verify that 
in the limit $E\searrow E_1$ we have $\mu\nearrow E_1$ with $T\searrow 0$
(or if $E\nearrow E_N\folgt \mu\searrow E_N$ with $T\nearrow 0$; 
see also Figs.~2,\ref{fig_mu_T_BE}). 
Also the micro-canonical expression for $\rho_m$ provides numerically 
correct $S(E)$ curves (identical to the grand canonical curve) 
for the lower 20\%-30\% of the energy spectrum where $\mu\approx E_1$ 
with a rather good accuracy. 

The Bose-Einstein ansatz with 
\lformula{\label{eq_BE}\rho_m=\rho_{BE}(E)\equiv \frac{1}{e^{(E_m-\mu)/T}-1}}
cannot be directly justified by the classical field approach. From a purely
formal point of view it can be obtained by replacing in the partition function 
$|C_m|^2\to c_m$ with integer $c_m$ and replacing 
the Gaussian integrations by sums over $c_m=0,1,2,\ldots$ thus 
resulting in (\ref{eq_BE}).
In the framework of this approach $T$ and $\mu$ are computed by solving 
numerically the implicit equations (\ref{eq_imp}) with $\rho_m=\rho_{BE}(E_m)$ 
which is technically a bit more complicated as for the EQ case. 
In the limit of large $|T|$ we can expand in (\ref{eq_BE}) the exponential 
and both approaches become equivalent. 

The difference between both approaches in the $S(E)$ curves is not very 
strong but the numerical data of long time averages of 
$\rho_m=\langle |C_m(t)|^2\rangle$ clearly show the validity of the EQ model 
provided the state is sufficiently thermalized as can be seen 
in Figs.~3, \ref{fig_states2}. 

Furthermore, according to both the micro-canonical and grand-canonical 
approaches the statistical distribution of $C_m$ is a complex Gaussian 
which corresponds to an exponential distribution of $|C_m|^2$, i.e. 
the distribution of the rescaled variable $x=(E_m-\mu)|C_m|^2/T$ 
is theoretically $p(x)=\exp(-x)$ which is clearly confirmed by the 
numerical data for quite large values of $x$ as can be seen in 
Fig.~\ref{fig_hist_m0_8} providing an additional confirmation 
of the classical model. 

Both approaches require the use of a given fixed energy spectrum $E_m$ 
which is typically obtained by diagonalizing a certain realisation of an 
RMT matrix (or another matrix for the variants as DANSE or the model 
with additional diagonal elements). However, in Fig.~1(d), we 
show the data for 10 different RMT realisations which would provide 
individually slightly different $S(E)$ curves. For this figure we used, 
for both theoretical $S(E)$ curves, 
a fictitious spectrum with $E_m$ being the solution 
of $m-1/2=M(E_m)$ for $m=1,\ldots, N$ where 
\formula{M(E)&=\frac{2N}{\pi}\int_0^E\sqrt{1-E'^2}\,dE'\\
&= \frac N2+\frac{N}{\pi}\left(\arcsin(E)+E\sqrt{1-E^2}\right)}
is the integrated density of states of the RMT semi-circle law 
such that $M(-1)=0$ and $M(1)=N$. This fictitious spectrum corresponds 
to a constant uniform level spacing in the unfolded spectrum.

The link between the radius (here being unity) 
of the semi-circle law of a GOE matrix and the variance of its 
matrix elements $\langle H_{n,n'}^2\rangle = (1+\delta_{n,n'})/(4(N+1))$ 
is rather standard [22]. However, it can be easily verified 
by computing the average 
\formula{\left\langle\mbox{Tr}(\hat H^2)\right\rangle=
\frac{1}{4(N+1)}\big(2N+N(N-1)\big)=\frac{N}{4}
}
which should coincide with 
\formula{\sum_m\left\langle E_m^2\right\rangle=
\frac{2N}{\pi}\int_{-1}^1 dE\, E^2\sqrt{1-E^2}=\frac{N}{4}\ .
}

The above derivations of the thermal distributions $\rho_{BE}$ and $\rho_{EQ}$
are done for finite size systems with a finite number of degrees of freedom.
However, they mainly follow the textbook approach of statistical physics 
for systems in the thermodynamical limit with an infinite number of 
degrees of freedom.

\section{Symplectic integrator}

Here we remind some basic facts about symplectic integrators 
and the particular implementation for our case. For further details, 
its derivation, we refer for example to [36,37,38], 
especially for the 4th order variant [36]. 
These kind of methods are also known as splitting methods [37,38]. 

\subsection{General method}

Let $A$ and $B$ be two non-commuting operators of a general Lie algebra 
for which it is possible to compute exactly and efficiently (by some 
exact numerical/analytical method) $\exp(tA)$ and $\exp(tB)$ individually and 
for arbitrary values of $t$ 
(or more precisely these operators applied to some given vector 
or function) while 
the numerical problem to compute $\exp[t(A+B)]$ is very difficult 
(very inefficient) or even impossible (as far as an exact method is concerned).

To solve this problem it is sufficient to compute 
$\exp[\Delta t(A+B)]$ for small $\Delta t$ (with some given precision) 
and then to apply: $\exp[t(A+B)]=\exp[\Delta t(A+B)]^n$ 
with $n=t/\Delta t$ (assuming that $t$ is an integer multiple of $\Delta t$).
To compute $\exp[\Delta t(A+B)]$ approximately one can write:
\formula{\exp[\Delta t(A+B)]\approx\prod_{j=1}^p \Big[\exp(d_j\Delta tA)\,
\exp(c_j\Delta tB)\Big]}
where the product is ordered with increasing $j$-values from right to left. 
The coefficients $c_j$, $d_j$, $j=1,\ldots, p$ are determined such that 
the error (for one step) 
is $\sim (\Delta t)^{p+1}$ for a given order $p$ 
and implying a global error $\sim (\Delta t)^p$ (for many steps and 
fixed $t$). 
The simplest case is $p=1$ with $c_1=d_1=1$ corresponding to the usual 
Trotter formula. For $p=2$, we have the symmetrized Trotter formula 
with $c_1=0$, $c_2=1$, $d_1=d_2=\frac12$. For $p=3$ there is a non-symmetric 
solution which can also be found in [36] 
(see references therein for the proper credit) but which is not really 
simpler (with all 6 coefficients being different from zero) than the fourth 
order solution. For $p=4$ there is a symmetric solution which according to 
[36] is:
\formula{c_1&=0,\ c_2=c_4=2x+1,\ c_3=-4x-1,\\
d_1&=d_4=x+0.5,\ d_2=d_3=-x}
where $x=(2^{1/3}+2^{-1/3}-1)/6$ is the real solution of 
$48 x^3 + 24 x^2 - 1 = 0$. Note that these coefficients verify the sum rule 
$\sum_j c_j=\sum_j d_j=1$ due to the first order terms in both 
exponential expressions. 
The fourth order formula requires as the third order formula the 
multiplication of 6 exponential factors for one step 
if one uses an optimization 
to merge the $d_4$-factor with the $d_1$-factor of the next step 
(a similar optimization is possible for the symmetrized Trotter formula). 

In typical applications one applies this method to solve numerically 
the time evolution of a classical Hamiltonian or a quantum system 
where the Hamiltonian is a sum of two terms $H_1+H_2$ for which the individual 
exponentials (of either the Liouville operator associated to $H_j$ or 
$-i H_j$, $j=1,2$) can be computed analytically or by an efficient 
exact numerical method. The splitting method can also be applied to 
a certain type of partial differential equations [37,38] with potential 
complications due to time steps of different signs (i.e. coefficients 
$c_j$ or $d_j$ having different signs). 
However, in our situation where we have a system 
of {\em ordinary differential equations} for a finite number of degrees 
of freedom, with an additional imaginary factor $i$ applied to the 
time variable, there is no numerical nor stability problem with respect 
to time steps of different signs. 

The advantage of the method is that it respects 
the symplectic/unitary symmetry of the problem. 
Furthermore, even if one chooses a low order variant with a not so small 
time step $\Delta t$, one can argue that the approximate time evolution 
(with respect to ``$A+B$'') represents in reality the {\em exact} time 
evolution of a slightly different operator $S\approx A+B$ such that 
$\exp(\Delta t S)$ coincides exactly with the above product of 
exponential terms and that many physical features of the modified time 
evolution are still very relevant since they apply to the same 
``class'' of systems. 

\subsection{Numerical implementation}

In our case, we chose $A=-i\hat H$ (in the quantum point of view 
or the Liouville operator associated to $\hat H$ in the classical point of  
view) and $B=-i\hat V(\psi)$ where $\hat V(\psi)$ is an effective potential 
depending on $\psi$ and with matrix elements 
$V_{n,n'}(\psi)=\beta |\psi_n|^2\,\delta_{n,n'}$. In this case 
$e^{-it\hat V(\psi)}$ provides the {\em exact} time evolution of the 
pure nonlinear equation (assuming $\hat H=0$): 
\formula{\frac{\partial\psi_n(t)}{\partial t}
&=-i\beta|\psi_n(t)|^2\,\psi_n(t)\folgt\\
\psi_n(t)&=e^{-it\beta|\psi_n(0)|^2}\,\psi_n(0)
}
which can be easily verified by writing $\psi_n=r_n\,e^{-i\theta_n}$ 
such that $\dot r_n=0\folgt r_n(t)=r_n(0)=\ $const. and 
$\dot\theta=\beta r_n^2\folgt \theta(t)=\theta(0)+t\beta r_n^2(0)$. 
The conservation of $|\psi_n(t)|=\ $const. (for the pure nonlinear 
equation) is a feature of the particular form of the nonlinear term 
and due to this $V(\psi)$ does not depend on 
$\psi$ nor on $t$ (during the purely 
nonlinear time evolution) and 
the time evolution due to the quantum exponential of $-it \hat V(\psi)$ 
coincides exactly with the time evolution of the exponential 
of the classical Liouville operator associated to the nonlinear term.

In the numerical implementation, we choose a certain initial 
condition of the state in the representation of the eigenbasis of $\hat H$, 
e.g. $C_m(0)=\delta_{m,m_0}$ with $m_0$ being the index of the 
initial state. Then, we apply the first exponential factor with coefficient 
$d_1$ (and given value of $\Delta t$) which corresponds to 
$e^{-iE_m d_1\Delta t}\,C_m\to C_m$. Then, using the unitary matrix 
that diagonalizes $\hat H$, we transform $C_m\to \psi_n$ and 
we apply the exponential factor with $c_2$ (since $c_1=0$ if 
$p=2$ or $p=4$) which corresponds to 
$e^{-ic_2\Delta t \beta|\psi_n|^2}\,\psi_n\to\psi_n$ 
which represents {\em exactly} the purely nonlinear time evolution. 
Then we transform $\psi_n\to C_m$ and apply the next exponential factor 
with coefficient $d_2$ etc. 
(If one uses a non-symmetric variant, with $c_1\neq 0$, for 
$p=1$ or $p=3$ one has 
first to transform the initial condition to $\psi_n$, apply the 
first $c_1$-factor and transform back to $C_m$.)

We have implemented and tested all four variants of the method. 
In particular, we have verified that the classical energy is conserved, i.e. 
its residual numerical fluctuations ($\sim 10^{-8}$ for the fourth order 
variant at $\Delta t=0.1$) scale with $(\Delta t)^p$ and also that 
the errors of other quantities scale with $(\Delta t)^p$.
For the case of a RMT with an extra diagonal where the values 
$E_m$ become larger, we have also tested the precision by comparing 
some data with $\Delta t=0.0125$ (for reduced iteration times) 
which does not change the values of $S$ etc. (apart from statistical 
fluctuations).

\newpage
\section{Additional Figures}
\vspace{0.5cm}
\noindent In this section, we present additional SupMat Figures 
for the main part of this article. The figure captions and figure notes 
contain physical discussions or additional information for figures in 
the main part; in particular the values of $T$ and $\mu$ for both 
approaches and the four states shown in Fig.~3 are given 
in the caption of Fig.~\ref{fig_states2} below.

\setcounter{figure}{0} \renewcommand{\thefigure}{S\arabic{figure}} 

FIGURE NOTES

\noindent {\bf Notes Fig.~\ref{figS1}:} The initial states are linear eigenstates 
$\phi_n^{(m_0)}$ of $\hat H$ (i.e. 
$\sum_{n'} H_{n,n'}\phi_{n'}^{(m_0)}=E_{m_0}\phi_{n}^{(m_0)}$) 
with specific values of $m_0$ given in the figure. The entropy is computed 
from $S(t)=-\sum_m \rho_m\ln(\rho_m)$ 
where $\rho_m$ is obtained as the time average 
$\rho_m=\langle |C_m(t)|^2\rangle$ 
for successive time intervals with increasing lengths by 
a factor of two corresponding to the plateau intervals of 
constant $S(t)$ visible in the figure. The thick horizontal lines represent 
the theoretical entropy $S_{EQ}$ for EQ (blue) and $S_{BE}$ for 
BE (red) for the energy of the state at $\beta=1$ and $m_0=3$ 
(pink open squares). 
At intermediate times $t\approx 2\times 10^4$ the entropy of 
this state is close to $S_{BE}$ while at longer times 
$t\ge 10^6$ it decreases to $S_{EQ}$ showing that the 
EQ ansatz describes the correct long time thermalization but 
also that at intermediate times the entropy is larger and closer 
to the BE ansatz. 
The other states $m_0=11,30,57$ at $\beta=1$ thermalize rather 
quickly at $t\ge 10^4$-$10^5$ to their final value $S_{EQ}$ 
(with $S_{BE}\approx S_{EQ}$ for $m_0=30$). 
For $\beta=0.1$ the state $m_0=30$ (cyan full squares) 
thermalizes to the same entropy 
value as with $\beta=1$ (green crosses) but only for very long time scales 
$t\ge 10^6$. 

\begin{figure}[H]
\begin{center}
\includegraphics[width=0.42\textwidth]{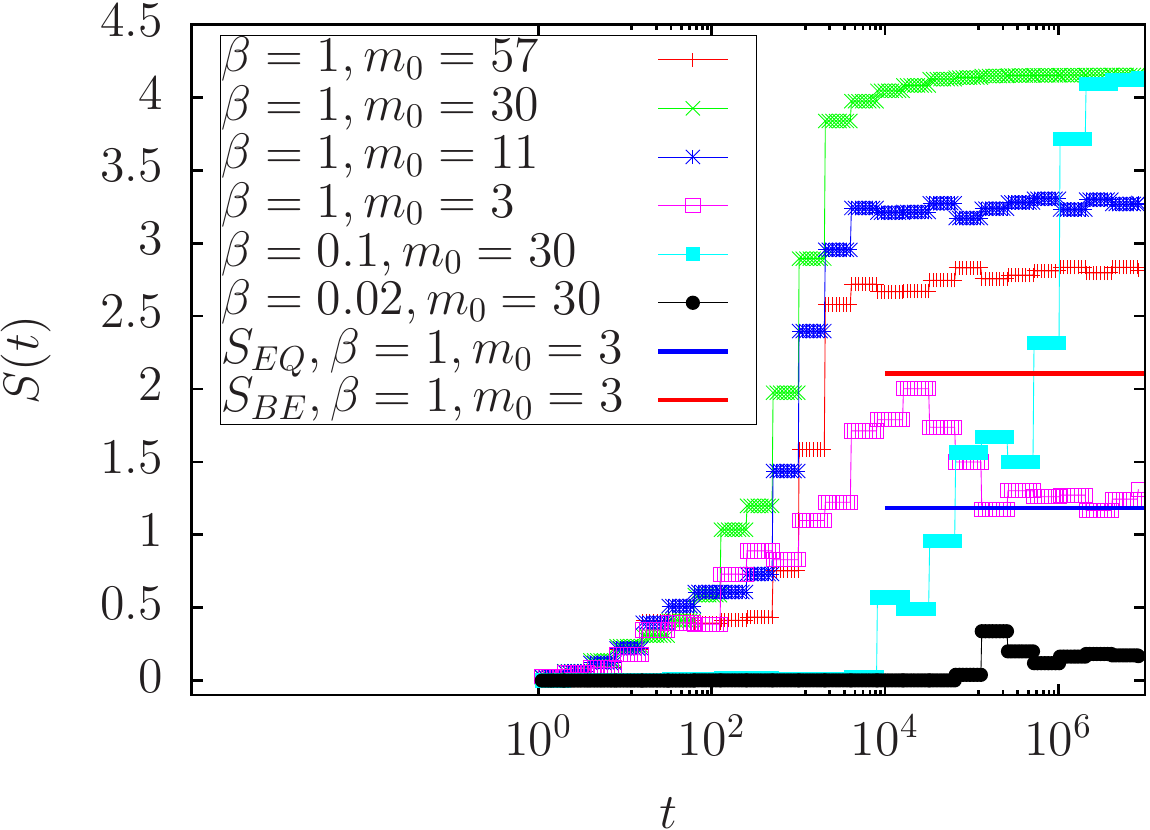}%
\end{center}
\caption{\label{figS1}
\label{fig_entropy_time}
Time dependence of the entropy $S(t)$ for the four states 
shown in Fig.~3 and two other states at $\beta=0.02,0.1$ 
with initial index $m_0=30$ (for $\beta=0.02,0.1$ the 
other three index values $m_0=3,11,57$ correspond to very small 
entropy $S(t)$ values clearly below the case $m_0=30$).
See details in FIGURE NOTES of Fig.~\ref{figS1}.
}
\end{figure}

\noindent {\bf Notes Fig.~\ref{figS2}:} For $\beta=1$, $N=256$ the time average corresponds to 
$2^{21}\le t\le 2^{22}$ (red $+$, all 256 initial conditions) 
and $2^{25}\le t\le 2^{26}$ (blue $\circ$, 35 initial conditions 
with $1\le m\le 35$). 
The curves represent the theoretical $S(E)$-curves from the EQ (blue) 
and BE (red) approaches using the exact spectrum of the used 
RMT realisation. The data point with $S>1$ for $N=32$, $\beta=0.02$ 
is not saturated and still increasing at the given maximal time $t=2^{27}$.
The data for $\beta=1$, $N=128$ and $\beta=1,2$, $N=256$ 
coincide very well with the EQ 
ansatz. Also for $N=32$ the EQ ansatz is more appropriate. Here the 
small differences to the theoretical EQ-curve are due to the fact that 
on the $x$-axis the initial energy $E_{m}$ is used and not the 
averaged linear energy $\langle E\rangle=\sum_{m'}E_{m'}\rho_{m'}$ 
using the long time average $\rho_{m'}$ and which is slightly 
different from $E_m$ due to the nonlinear term. Using $\langle E\rangle$ the 
data points (for the cases with good thermalization) fall nearly exactly 
on the theoretical curve. 
For $\beta=1$, $N=256$ the data points with $m\ge 10$ thermalize well 
and rather {\em early} to 
the EQ curve already for the time average interval $2^{21}\le t\le 2^{22}$.
The data points for $1\le m\le 5$ and $m=7$ do not thermalize at all even 
for $2^{25}\le t\le 2^{26}$ with entropy values clearly below the EQ and 
BE curves and being rather constant between $2^{22}\le t\le 2^{26}$. 
The two data points for $m=6$ and $m=9$ thermalize {\em late} to the EQ 
curve for $2^{25}\le t\le 2^{26}$ while for $2^{21}\le t\le 2^{22}$ their 
entropy values are clearly below the EQ and BE curves. The data 
point at $m=8$ also approaches {\em late} the EQ curve 
($2^{25}\le t\le 2^{26}$) but from above, i.e. 
with {\em early} entropy values ($2^{21}\le t\le 2^{22}$) 
slightly above the EQ curve but still clearly 
below the BE curve. Additional and more detailed figures (higher 
resolution and more data points at different times) for theses 
points are available at [39].

We note that for $N=128,256$ detains states close to 
the spectral border $E = \pm 1$ are not thermalized even
at very large times. We attribute this to the fact that at such energies
the energy level spacing is significantly increased as 
compared to the band center and thus a stronger nonlinearity $\beta$ is 
required for thermalization. 
Indeed, for the larger nonlinearity parameter 
$\beta=2$ more states of those border states 
are thermalized as compared to $\beta=1$. 
More data for $N=256$ are available in [39].

\begin{figure}[H]
\begin{center}
  \includegraphics[width=0.42\textwidth]{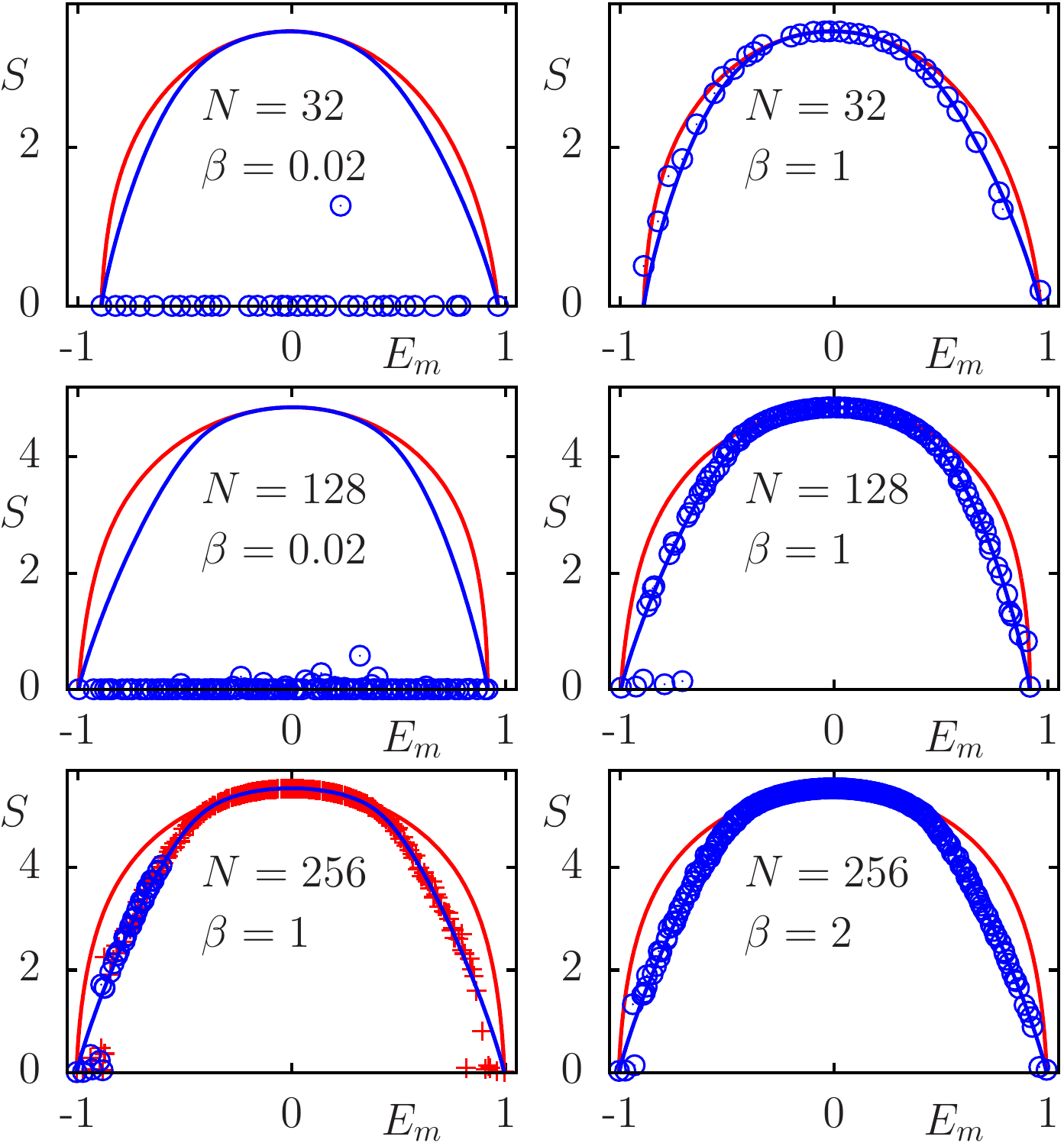}%
\end{center}
\caption{\label{figS2}
\label{fig_SE2} Dependence of entropy on energy $S(E)$.
As in Fig.~1 data are for one RMT realisation, $\beta=0.02,1$ for $N=32,128$ 
and $\beta=1,2$ for $N=256$, 
$\rho_m$ obtained by the time average for $2^{21}\le t\le 2^{22}$ for 
$N=128,256$ and $2^{26}\le t\le 2^{27}$ for $N=32$ (blue $\circ$, 
all panels except $\beta=1$, $N=256$).
See details in FIGURE NOTES of Fig.~\ref{figS2}.
}
\end{figure}

\newpage
\noindent {\bf Notes Fig.~\ref{figS3}:} The data points in Figs.~2 and \ref{figS3} 
were obtained by computing $\mu$ and $T$ from 
the implicit set of the two 
equations~: $S_{\rm num.}=-\sum_m \rho_m\ln\rho_m$ and 
$1=\sum_m \rho_m$ using the expressions Eq.~(2) for both 
approaches and the numerical values of $S_{\rm num.}$ 
(data points in Figs.~1 and \ref{figS2}). Therefore, 
if $S_{\rm num.}$ and $S(E)$ are not identical (due to 
statistical fluctuations or lack of thermalization), these data points 
do not need to fall exactly on the theory curves which were obtained 
by solving another set of two equations~: 
$E=\sum_m E_m\rho_m$ and $1=\sum_m\rho_m$ (using Eq.~(2)).
The deviations of the data points with respect to the theory curves 
are significantly weaker for the EQ case than for the BE case but 
the latter are still quite weak, even though better 
visible in Fig.~\ref{figS3} as compared to Fig. 2 (with no visible difference 
on graphical precision). 
This observation confirms somehow that the EQ ansatz fits better the 
numerical data but the analysis shown in Figs.~2 and \ref{figS3} does 
not allow to distinguish very clearly between the validity 
of either the EQ or the BE ansatz. 

For this the direct comparison the numerical values of $\rho_m$ 
with the expressions (2) (see Figs. 3 and \ref{figS4}), provide 
a much stronger argument in support of the EQ ansatz. 

\begin{figure}[H]
\begin{center}
\includegraphics[width=0.42\textwidth]{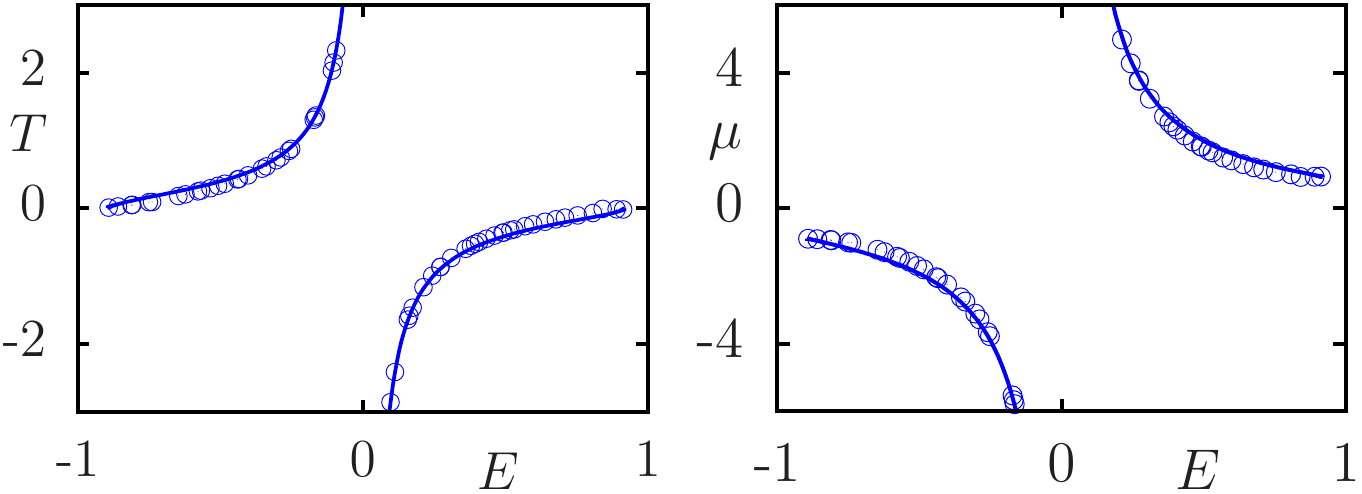}%
\end{center}
\caption{\label{figS3}
\label{fig_mu_T_BE}
As Fig.~2 but for the BE case. Here the 
data points for $T$ and $\mu$, computed from the numerical 
data of $S$, show some small deviations from the theoretical curves 
which are visible in the figure and significantly larger than 
in Fig.~2 where no deviations for the EQ case 
(on graphical precision) are visible.
Furthermore, in comparison to the EQ case of Fig.~2
the typical values of $T$ and $\mu$ for the BE case are considerably 
larger. See details in FIGURE NOTES of Fig.~\ref{figS3}.
}
\end{figure}

\noindent {\bf Notes Fig.~\ref{figS4}:} Here, the blue curve shows the theoretical 
values based on EQ 
with $\rho_{\rm EQ}(E)=T/(E-\mu)$ and
$T=0.002007,0.004986$ ($T=0.0006853, 0.001414$), 
$\mu=-0.9995, -1.041$ ($\mu=-1.006,-1.009$) 
for $m_0=11,31$ and $N=128$ ($N=256$). 
$T$ and $\mu$ were determined from the solution of the 
equations $1=\sum_m \rho_{\rm EQ}(E_m)$ and 
$\langle E\rangle=\sum_m E_m\rho_{\rm EQ}(E_m)$
with $\langle E\rangle=\sum_m E_m \rho_m \approx E_{m_0}$.
The red line shows the theoretical 
values based on BE  
with $\rho_{\rm BE}(E)=1/(\exp[(E-\mu)/T]-1)$, 
$T=0.1983,0.5562$ ($T=0.1301,0.2686$),
$\mu=-1.434,-2.914$ ($\mu=-1.300,-1.875$) 
for $m_0=11,31$ and $N=128$ ($N=256$). 
Here $T$, $\mu$ were determined from the solution of the 
equations $1=\sum_m \rho_{\rm BE}(E_m)$ and 
$\langle E\rangle=\sum_m E_m\rho_{\rm BE}(E_m)$.
Furthermore, the energy values for $m_0=11,31$ and $N=128$ ($N=256$) 
are $E_{m_0}\approx\langle E\rangle=-0.7426,-0.4031$ 
($E_{m_0}\approx\langle E\rangle=-0.8307,-0.6469$). 
The thermalization of all four states according to the EQ theory 
is very good (essentially perfect) 
despite the shorter averaging time for the case $N=128$
as compared to Fig.~3. For $N=256$ the averaging time 
is rather long but here the selected states are closer to the band edge 
at $E_{\rm min}=-1$ and with lower temperature values (than for $N=128$) 
such that thermalization is more difficult.
Additional similar figures for other values of $m_0$ at different values of 
$N$ are available at [39]. 

These results clearly show that the dynamical thermalization
of $\rho_m$ is very well described by the EQ ansatz (2).
High quality figures of $\rho_m$ for all initial states values $m_0$ 
at $N=256$ are available in [39],
including all thermalized states with negative temperatures $T <0$
which appear at energies  $E_{m_0} > 0$.

\begin{figure}[H]
\begin{center}
\includegraphics[width=0.42\textwidth]{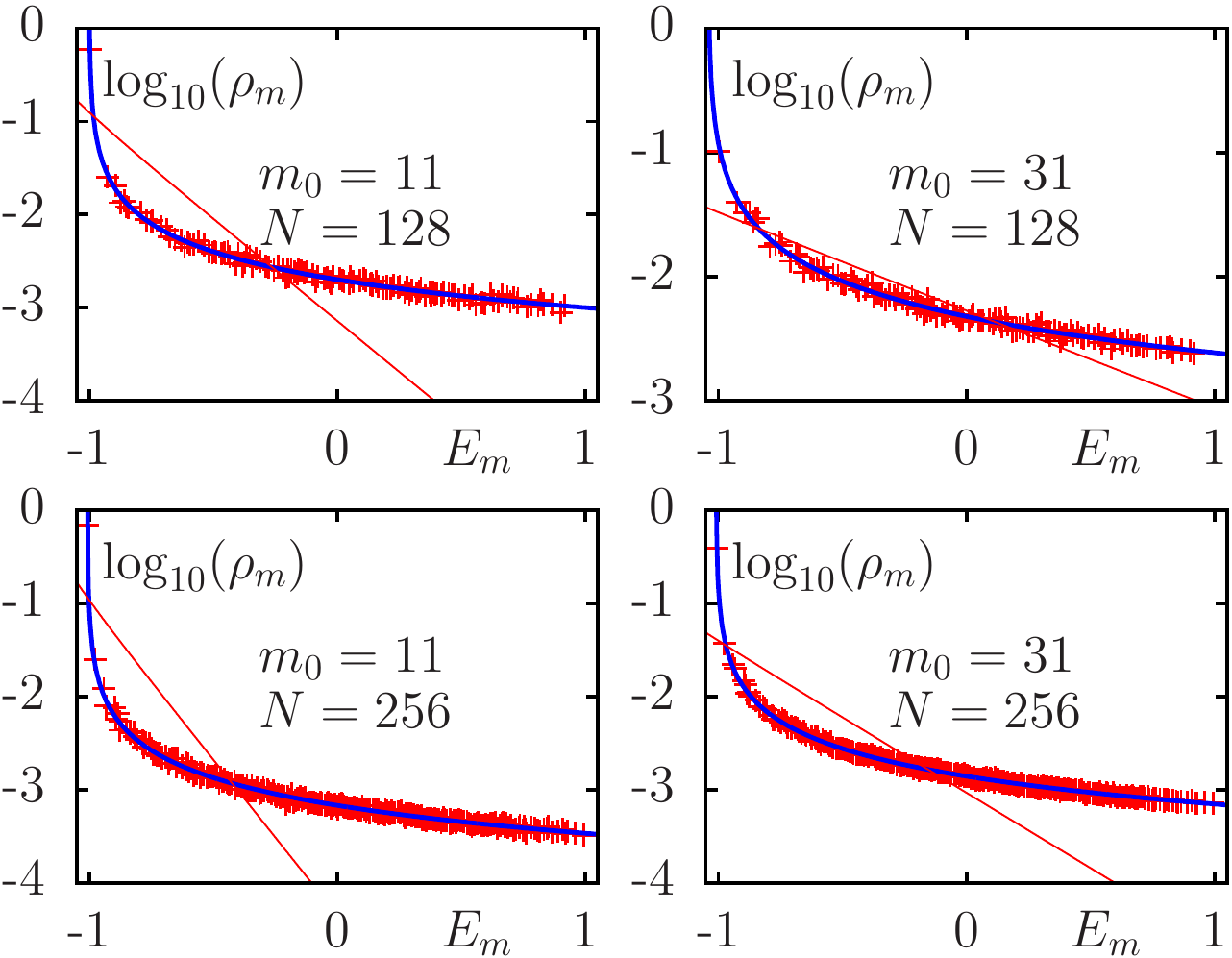}%
\end{center}
\caption{\label{figS4}
\label{fig_states2}
Dependence of $\rho_m$ on $E_m$ for two states with initial 
state $m_0=11,31$ for $\beta=1$, $N=128$ ($N=256$) obtained by an 
time average in the interval $2^{21}\le t\le 2^{22}$ 
($2^{25}\le t\le 2^{26}$). 
{\it Complementary information for} Fig.~3:
The $T$ and $\mu$ values for the EQ ansatz and the four states 
$m_0=3,11,30,56$ with $N=64$ and $\beta=1$ of Fig.~3 are 
$T=0.001372,0.005984,0.07585,-0.004538$, 
$\mu=-0.8964,-0.9178,-4.892,0.9293$ 
and the corresponding values for the same states and the BE ansatz 
are $T=0.112,0.3581,4.913,-0.2649$, $\mu=-1.062$, $-1.794$, $-20.52$, $1.496$.
See details in FIGURE NOTES of Fig.~\ref{figS4}.
}
\end{figure}

\newpage
\noindent {\bf Notes Fig.~\ref{figS5}:} This figure clearly shows 
that the statistical distribution of $C_m(t)$ 
(or of the quantity $u\equiv |C_m(t)|^2$) is very well described by the 
thermal Boltzmann Gaussian distribution $\exp(-(E_m-\mu)|C_m(t)|^2/T)$ 
(or exponential distribution $\exp(-(E_m-\mu)\,u/T)$
in $u$) for values up to $u\approx (8$-$10)\times\langle u\rangle$.
\begin{figure}[H]
\begin{center}
\includegraphics[width=0.42\textwidth]{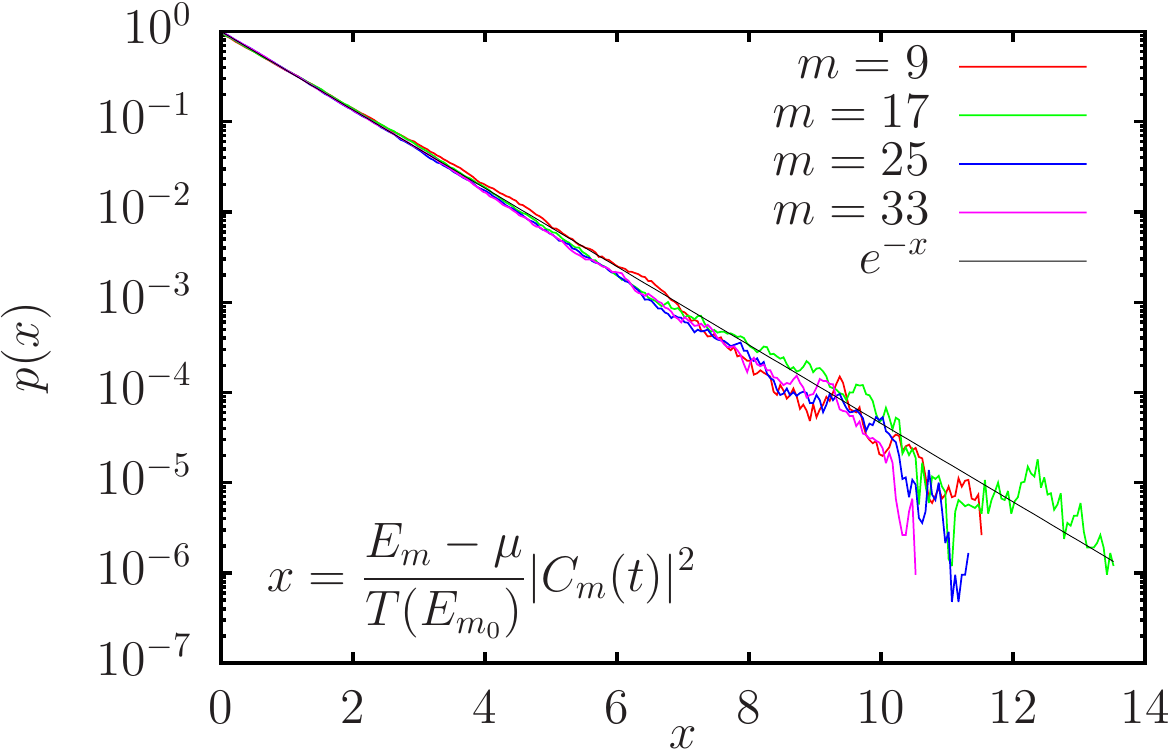}%
\end{center}
\caption{\label{figS5}
\label{fig_hist_m0_8}
Statistical distribution of the rescaled variable 
$x=(E_m-\mu)|C_m(t)|^2/T(E_{m_0})$ 
for $\beta=1$, $N=64$, $m_0=9$, $m=9,17,25,33$ 
using the time values in the interval 
$2^{23}\le t\le 2^{24}$ and a histogram of bin width 0.05.
The thin black line shows the theoretical distribution $p(x)=e^{-x}$ 
according to the EQ approach. 
The numerical distributions follow the theoretical distribution 
for values up to $x\approx 8$-$10$ providing an additional confirmation 
for the validity of the EQ ansatz.
See details in FIGURE NOTES of Fig.~\ref{figS5}.
}
\end{figure}

\noindent {\bf Notes Fig.~\ref{figS6}:} This figure is similar to Fig.~5 but with 
additional $\beta$ values~:
Lyapunov exponent $\lambda_{m}$ dependence on $E_{m}$ with 
$m$ being the index of the initial state $\phi^{(m)}$ for $N=64$.
Here $\lambda_{m}$ 
is determined from the fit
$\ln\|\Delta\psi(t)\|=a+b\ln(t)+\lambda_{m}\,t$ for $t\le 2^{22}$ 
and $\beta=2$ (grey $\blacktriangle$; top),
$\beta=1.5$ (orange $\vartriangle$),
$\beta=1$ (black $\bullet$),
$\beta=0.75$ (cyan $\blacksquare$),
$\beta=0.5$ (pink $\square$),
$\beta=0.25$ (blue $*$),
$\beta=0.1$ (green $\times$),
$\beta=0.02$ (red $+$),
$\beta=0.02$ for $t\le 2^{27}$ (red $\triangledown$; bottom).
The numerical data suggests that most $\lambda_m$ for $\beta=0.02$ 
decay as $\lambda_m\sim 1/\sqrt{t}$ for $t\ge 10^7$ (see 
Fig.~\ref{fig_lyapunov_time2} below). However, 
three $\lambda_m$ values for $\beta=0.02$ do not decay with 
time (data points with same red $+$ and $\triangledown$; e.g. 
$m=45$ and $E_m\approx 0.35$) and 
have significantly larger values $\lambda_m>10^{-5}$ indicating a 
trajectory in a chaotic region while other initial conditions correspond 
to trajectories in bounded KAM regions. These cases are also visible 
in Fig.~1 (a) with entropy values slightly above 0.

\begin{figure}[H]
\begin{center}
\includegraphics[width=0.42\textwidth]{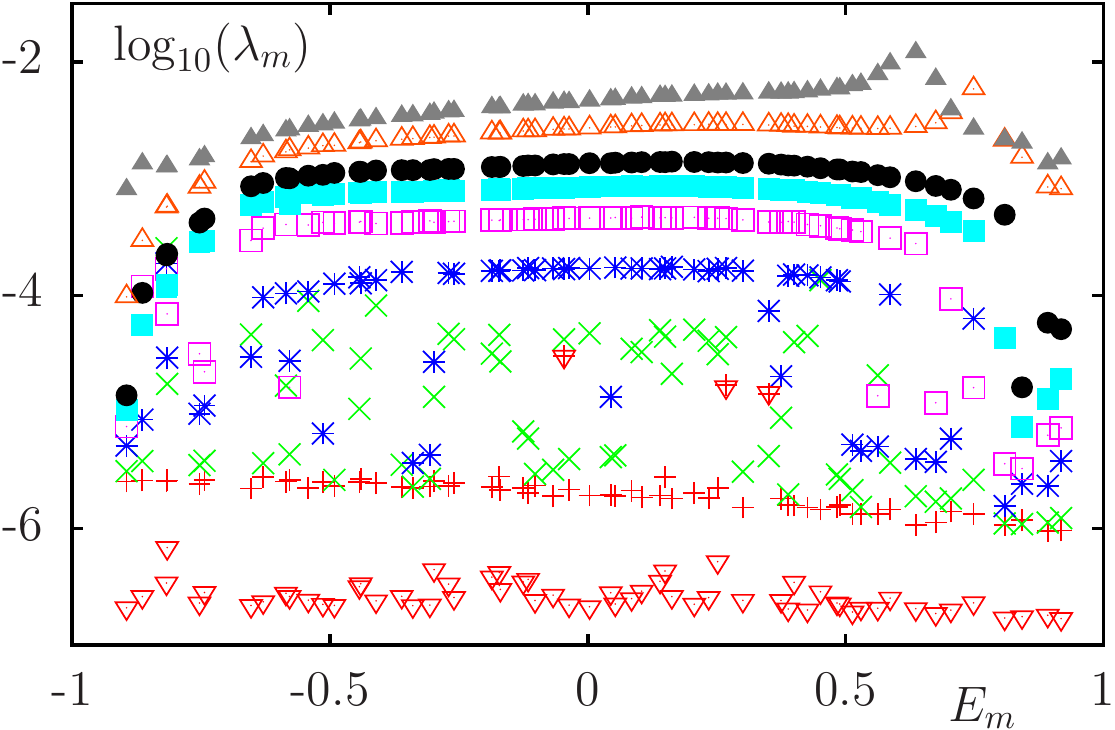}%
\end{center}
\caption{\label{figS6}
  \label{fig_lyapunov_logS1} Dependence of Lyapunov exponent $\lambda_m$
  on the energy  $E_m$ of the initial state $\phi^{(m)}$, same as in  Fig.~5.
See details in FIGURE NOTES of Fig.~\ref{figS6}.
}
\end{figure}

\newpage
\noindent {\bf Notes Fig.~\ref{figS7}:} In the {\it bottom} panel
   the fits are done using all data points such that the values 
for $\beta\le 0.05$ have a smaller weight due to the reduced number 
of different initial conditions. Very long time computations 
for 35 random initial conditions at $\beta=0.02$ for $t\le 3\times 10^9$ 
($t\le 10^9$) for $N=32$ ($N=64$) indicate a chaotic behavior with 
no further global decrease of $\lambda$ with time 
for $t>2^{27}\approx 1.3\times 10^8$. There are however considerable 
fluctuations between different initial conditions and 
in the time dependence (but with very long correlation times) 
of the order of 10-15\%. Additional figures for this point 
are available at [39].

We point out 
that for a localized initial condition with only one mode $m$ 
the Lyapunov exponent at $\beta=0.02$ ({\it top} panel)
decreases with time as $\lambda\sim t^{-1/2}$ (see Fig. \ref{figS10})
indicating a non-chaotic behavior in the limit of very large times.
In contrast for states  $\psi(t) = \sum_m C_m(t) \phi^{(m)}$,
with uniform random initial $C_m$ amplitudes at $t=0$
({\it bottom} panel), the Lyapunov exponents
are well stabilized at large times $t < 1.3 \times 10^8$
even for $\beta = 0.02$ (see Figs. in [39]). 
These state have automatically an average energy 
$E\approx 0$ close to the band-center. 
It is important to stress that all states with initial random
configurations have approximately 
the same values of $\lambda >0$. This means that
even at small value $\beta =0.02$ the measure of the 
chaotic component (at $E \approx 0$) is close to unity.
This result is very different from
many-body nonlinear systems studied in [40]
where the measure of the chaotic component is 
$\sim\beta$.

\begin{figure}[H]
\begin{center}
\includegraphics[width=0.42\textwidth]{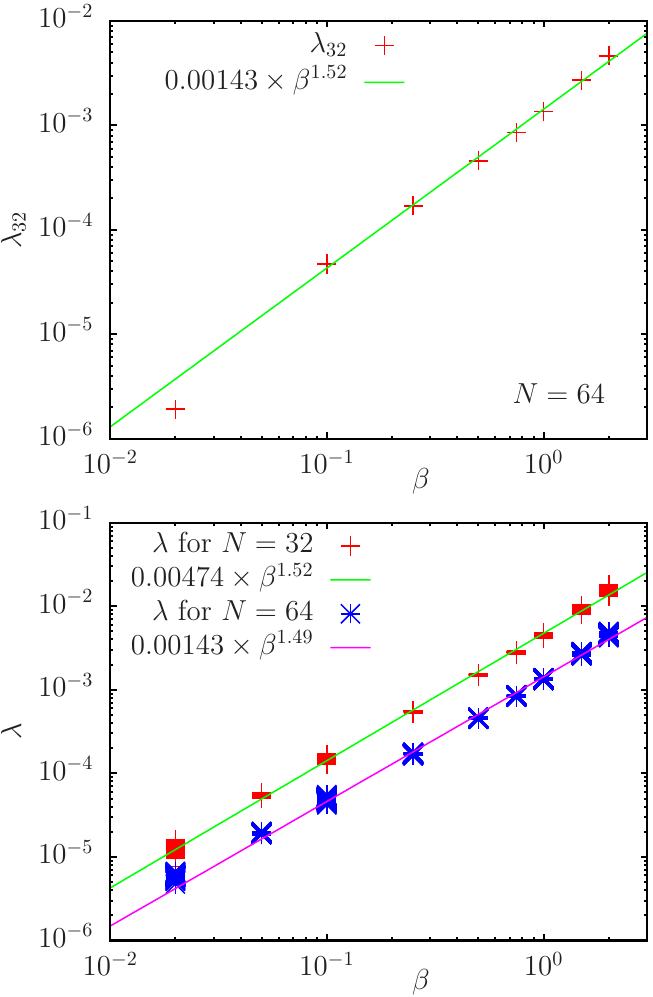}%
\end{center}
\caption{\label{figS7}
\label{fig_lambda_scale_beta_31}
      {\em Top:} Dependence of Lyapunov exponent $\lambda_{32}$ on $\beta$
      in the energy band center 
for the localized initial condition at $m=32$, 
$N=64$ and computation time $t=2^{22}$. 
The straight green line shows the power law fit 
$\lambda_{32}=a\beta^\eta$ with $a=0.00143\pm 0.00005$ and 
$\eta=1.52\pm 0.03$. For this fit the smallest data point at $\beta=0.02$ 
was not used since for this value the Lyapunov exponent continues to 
decrease with increasing computation time $t$ and it is most likely 
below the chaos border.
{\em Bottom:} Dependence of Lyapunov exponent $\lambda$ 
on $\beta$ for many different random uniform $C_m$ configurations
(most energy values $\langle E\rangle\approx 0$ and 
some cases with $\langle E\rangle\approx \pm 0.1$)
for $N=32,64$ and computation time $t=2^{22}$ ($t=2^{27}$) for 
$\beta\ge 0.1$ ($\beta\le 0.05$). 
The number of initial conditions is 64 for $\beta\le 0.05$, 
and (at least) $10N=320,640$ for $\beta\ge 0.1$;
symbols at fixed $\beta$ mark $\lambda$
for different initial configurations. 
The straight green line shows the power law fit 
$\lambda=a\beta^\eta$ with $a=0.00474\pm 0.000009$ and 
$\eta=1.524\pm 0.0015$ for the case $N=32$.
The straight pink line shows the power law fit 
$\lambda=a\beta^\eta$ with $a=0.001426\pm 0.000002$ and 
$\eta=1.491\pm 0.0013$ for the case $N=64$.
See details in FIGURE NOTES of Fig.~\ref{figS7}.
}
\end{figure}

\newpage
    \noindent {\bf Notes Fig.~\ref{figS8}:}
    In the {\it bottom} panel symbols at fixed $N$ show $\lambda_m$ values
    for different initial condition (one mode $\phi^{(m)}$ with energy
    close to the band center); all obtained $\lambda_m$ values
    are rather close to each other. This indicates
    that at $\beta=1$ the measure of the chaotic component is close to unity
    for $N \leq 512$ (see also NOTES of Fig.~\ref{figS7}).

\begin{figure}[H]
\begin{center}
\includegraphics[width=0.42\textwidth]{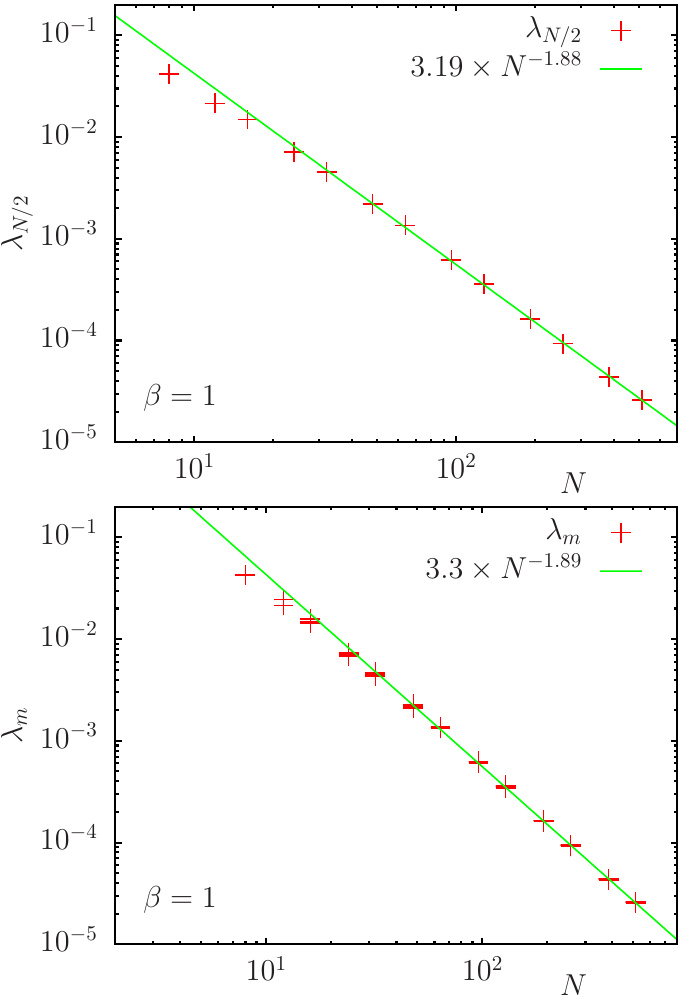}%
\end{center}
\caption{\label{figS8}
\label{fig_lambda_scale_N_31}
      Dependence of Lyapunov exponent $\lambda_{N/2}$ on $N$ at $\beta=1$.
      {\em Top:}  data for the band center 
      and computation time $t=2^{22}$ ($t=2^{24}$) 
for $N\le 384$ ($N=512$).  
The straight green line shows the power law fit 
$\lambda_{N/2}=aN^{-\nu}$ with $a=3.19\pm 0.19$ and 
$\nu=1.88\pm 0.01$ using the data for $N\ge 32$. 
{\em Bottom:} Same as top but showing
 all $\lambda_m$ values corresponding to 
$|E_m|\le 0.1$ (for $N\le 128$) or 
35 values in the band center with $|m-N/2|\le 17$ (for $N\ge 192$). 
The straight green line shows the power law fit 
$\lambda_{N/2}=aN^{-\nu}$ with $a=3.30\pm 0.04$ and 
$\nu=1.886\pm 0.002$ using the data for $N\ge 32$. 
The data of both panels correspond to the case of localized initial 
conditions at some value $m$ (with $E_m$ being close to the band center).
See details in FIGURE NOTES of Fig.~\ref{figS8}.
}
\end{figure}

\noindent {\bf Notes Fig.~\ref{figS9}:} At $\beta=1$ the scaling $\lambda \propto 1/N^\nu$
with $\nu=1.89$ works well in the energy band center. Certain deviations
from this  scaling are seen in the vicinity of the 
energy edges $ E \approx \pm 1$. We attribute this to
a significant increase of level spacing at band edges that
may modify chaos properties at different $N$ at band edges.

\begin{figure}[H]
\begin{center}
\includegraphics[width=0.42\textwidth]{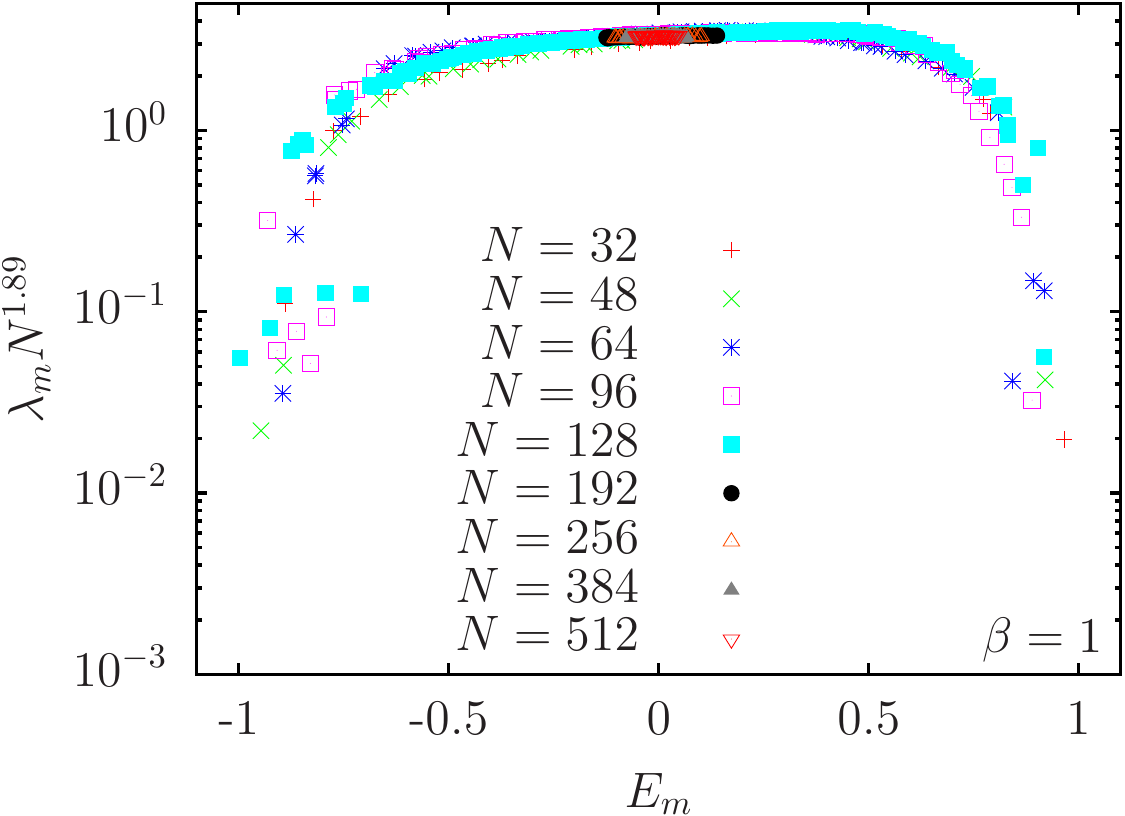}%
\end{center}
\caption{\label{figS9}
\label{fig_lyapunov_logNscale}
Dependence of rescaled Lyapunov exponent $\lambda_m N^\nu$ on 
the initial energy $E_m$ using the exponent 
$\nu\approx 1.89$ found in the fit of (the bottom panel of) 
Fig.~\ref{fig_lambda_scale_N_31} for different values of $N$ with 
$32\le N\le 512$ and $\beta=1$. 
For $N\ge 192$ only 35 values of $\lambda_m$ (per $N$ value) 
in the band center with $|m-N/2|\le 17$ are available and shown.
See also FIGURE NOTES of Fig.~\ref{figS9}.
}
\end{figure}

\newpage
\noindent {\bf Notes Fig.~\ref{figS10}:} Here 
$\lambda_m(t)$ has been obtained by the fit 
$\ln\|\Delta\psi(\tau)\|
=a+b\ln(\tau)+\lambda_{m}\,\tau$ for $0\le \tau\le t$ and for 
values $t\le 2^{27}$ 
where $\Delta\psi(\tau)=\tilde\psi(\tau)-\psi(\tau)$ is 
the difference vector between two close initial conditions with 
$\psi(0)=\phi^{(m)}$, $\tilde \psi(0)=\psi(0)+\Delta\psi(0)$ 
and $\Delta\psi(0)$ being a random vector with initial 
norm $\|\Delta\psi(0)\|=10^{-12}$. 
During the computation the difference vector $\Delta\psi(\tau)$ 
is regularly renormalized to the norm $10^{-12}$ when its norm has become 
larger than $10^{-10}$ such that both trajectories stay close and 
the logarithm of the renormalization factor is added to a special 
variable measuring the quantity $\ln\|\Delta\psi(\tau)\|$ which 
is used for the computation of the Lyapunov exponent. 
The rescaled logarithmic growth $(\ln\|\Delta\psi(t)\|)/t$ shows 
roughly the same behavior as $\lambda_m(t)$, with a final slope 
somewhat closer to the exponent $-1/2$ than for $\lambda_m(t)$ 
(in logarithmic representation and for $t\ge 10^7$).

The two cases at $m=31$, $r=0,1$
indicate a vanishing Lyapunov exponent in the limit $t\to\infty$ 
and a trajectory in a bounded KAM region. The Lyapunov exponent for 
the other two cases at $m=45$ (with $E_{45}(r=0)\approx 0.351$ 
and $E_{45}(r=1)\approx 0.310$) saturate to the values 
$\lambda_{45}(r=0)\approx 1.47\times 10^{-5}$ and 
$\lambda_{45}(r=1)\approx 8.89\times 10^{-6}$ 
in the limit $t\to\infty$ 
indicating a trajectory in a chaotic region probably due to the effect 
of a near triple quasi-resonance for the given RMT realisation. 
For the first realisation $r=0$ there are three cases like this 
as can be seen in Figs.~5 and \ref{fig_lyapunov_logS1} (see 
also caption therein). The observation that for both realisations 
there are saturated Lyapunov values at the same index $m=45$ is a coincidence 
and for example for $m=32$ (not shown in the figure) 
there is a stabilized Lyapunov exponent for $r=1$ but not for $r=0$. 

\begin{figure}[H]
\begin{center}
\includegraphics[width=0.42\textwidth]{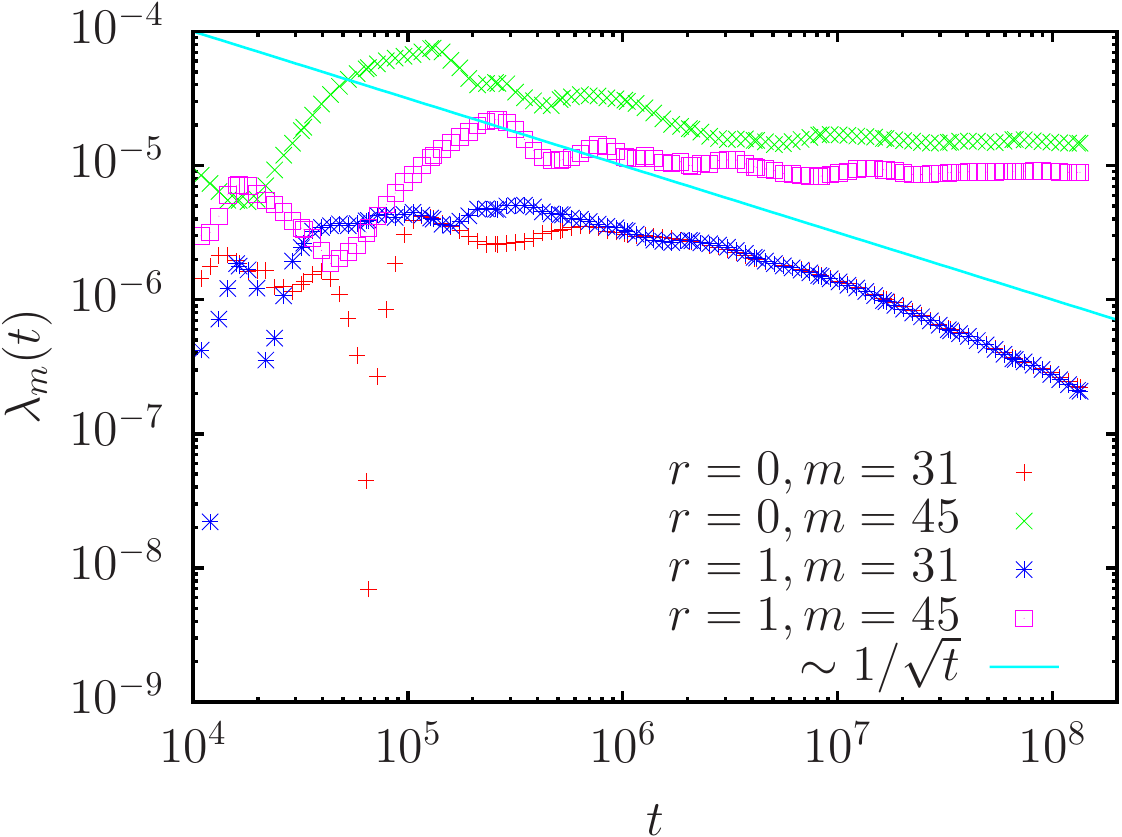}%
\end{center}
\caption{\label{figS10}
\label{fig_lyapunov_time2}
Dependence of Lyapunov exponent $\lambda_m(t)$ on  time 
$t$ for  $\beta=0.02$, $N=64$ and two initial states in the 
band center, $m=31,45$. Data are for two RMT realisations $r=0$ (same realisation 
as for most main and SupMat figures concerning the RMT case) and $r=1$. 
The cyan full line shows $10^{-2}/\sqrt{t}$ to indicate an empirical  $t^{-1/2}$ 
power law $\lambda \propto t^{-1/2}$ at large times.
See details in FIGURE NOTES of Fig.~\ref{figS10}.
}
\end{figure}

\noindent {\bf Notes Fig.~\ref{figS11} and Fig.~\ref{figS12}:}
These two figure correspond to the case of the DANSE model
studied in [29]. In the limit $N \rightarrow \infty$
and $\beta =0$ the model is reduced to the Anderson model
in one-dimensions with exponentially localized eigenstates
and the localization length $\ell \approx 96/W^2$
in the band center. Here $W$ is the strength of the diagonal disorder.
For $N=32, 64$ and $W=2, 4$ the value of $\ell$ is
comparable to the system size and chaos induced
by the nonlinearity $\beta$ leads to dynamical thermalization EQ (2)
as it is shown in Fig.~\ref{figS11} and Fig.~\ref{figS12}.
More thermalization figures for all initial eigenmodes,
including those leading to negative temperature $T<0$ are
available at [39].

\begin{figure}[H]
\begin{center}
\includegraphics[width=0.42\textwidth]{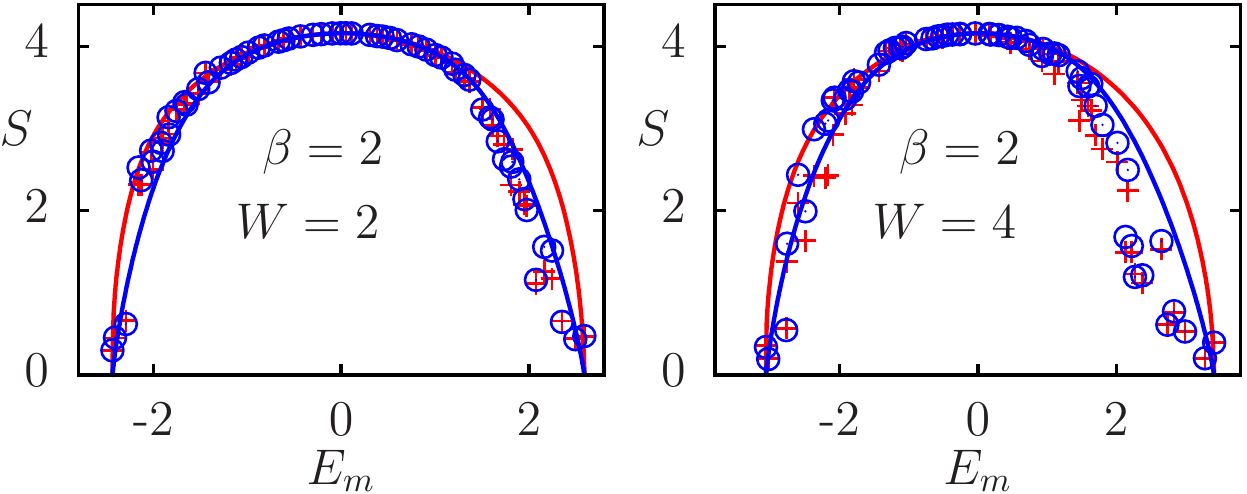}%
\end{center}
\caption{\label{figS11}
\label{fig_DANSE_SE}
As Fig.~1 but for one realisation of the DANSE model 
of [29] at disorder strength $W=2$ and $W=4$ for $\beta=2$ and 
$N=64$. The data points correspond to the averaging time 
$2^{23}\le t\le 2^{24}$ (blue $\circ$) and 
$2^{20}\le t\le 2^{21}$ (red $+$; similar $t$ values as in [29]).
See also FIGURE NOTES of Fig.~\ref{figS11} and Fig.~\ref{figS12}.
}
\end{figure}

\begin{figure}[H]
\begin{center}
\includegraphics[width=0.42\textwidth]{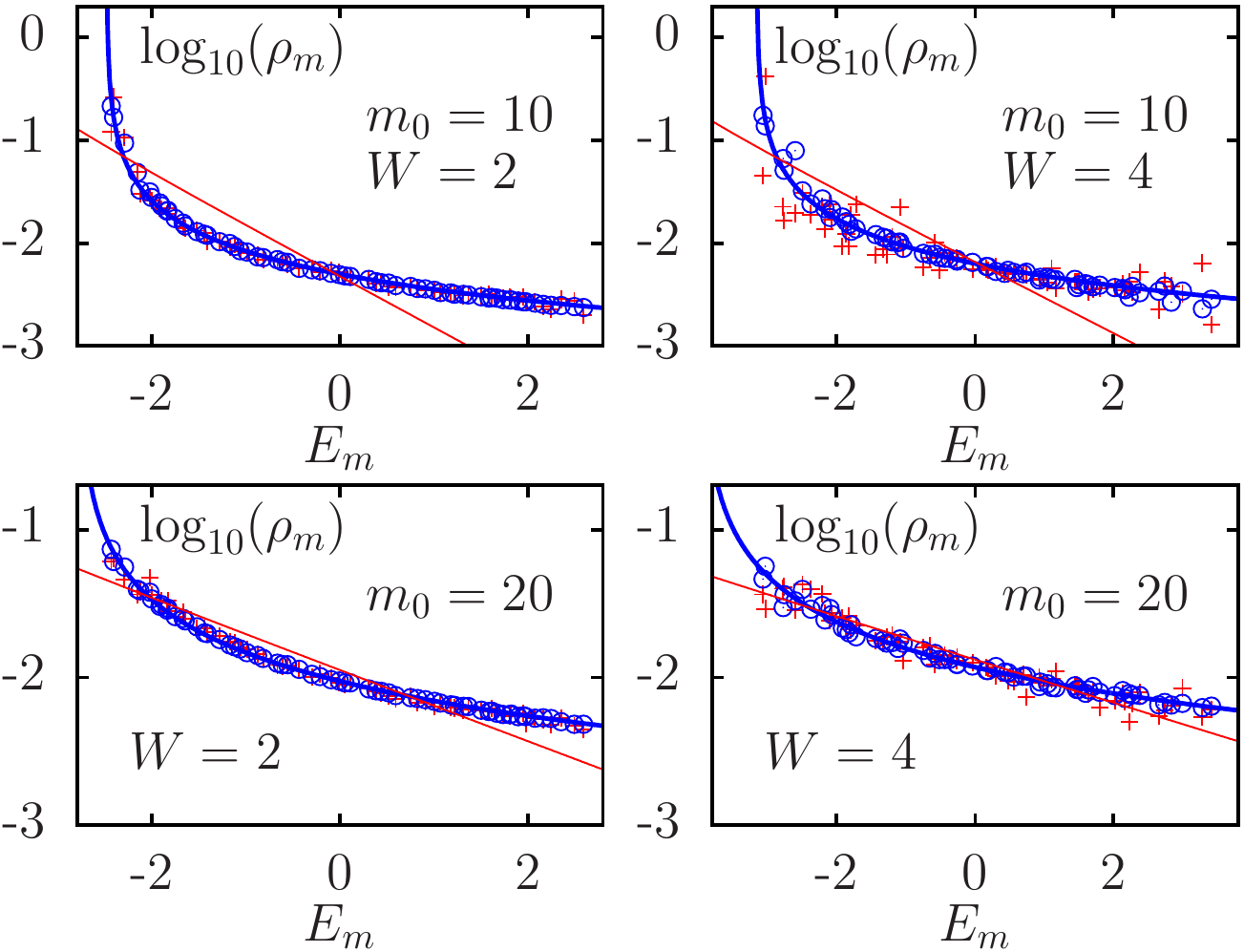}%
\end{center}
\caption{\label{figS12}
\label{fig_DANSE_state}
As Figs.~3 and \ref{fig_states2} but for one realisation of the 
DANSE model 
of [29] at disorder strengths $W=2$ and $W=4$ for $\beta=2$, $N=64$
and two initial states with $m_0=10,20$. 
The data points correspond to $\rho_m$ obtained by the averaging time 
$2^{23}\le t\le 2^{24}$ (blue $\circ$) and 
$2^{20}\le t\le 2^{21}$ (red $+$; similar $t$ values as in [29]).
The values of $T$ and $\mu$ for the EQ approach are 
$T=0.0124,0.02626,0.01997,0.04636$ and
$\mu=-2.484,-2.784,-3.157,-3.952$ for $W=2$ with $m_0=10,20$ and 
$W=4$ with $m_0=10,20$.
The values of $T$ and $\mu$ for the BE approach are 
$T=0.8794,1.815,1.26,3.011$ and 
$\mu=-4.709,-8.171,-6.346,-13.07$ for the same states.
For $W=2$ both states are well thermalized according to the EQ case. 
For $W=4$ the thermalization also corresponds to the EQ case but there 
are still stronger fluctuations, especially for the data with shorter 
averaging time (and corresponding to the data of [29]).
See also FIGURE NOTES of Fig.~\ref{figS11} and Fig.~\ref{figS12}.
}
\end{figure}

\noindent {\bf Notes Fig.~\ref{figS13} and Fig.~\ref{figS14}:} These figures
correspond to the case with an additional linearly growing
term $fn$ added to the diagonal matrix elements $H_{n,n}$.
At $\beta=1$, $N=32$  and $f=0.25$ the dynamical thermalization
is reached at large times
but it is not completely the case for $f=0.5$ (see Fig.~\ref{figS13}).
As shows Fig.~\ref{figS14} at initial times we have
an approximate exponential drop of probabilities $\rho_m$ with $E_m$
(red crosses) which is similar to a BE or quantum Gibbs distribution.
However, at larger times the distribution $\rho_m$
approaches the theoretical EQ curve (2). We assume that
there is a relatively rapid process of chaotic mixing of
modes being close to the initial $m=m_0$ value
and those with lower energies at $m < m_0$.
Somehow it is easy to go to low energies
while the propagation of excitations to higher energies, with
$m$ being significantly higher than $m_0$,
goes as a slow diffusion requiring significantly longer times.
Indeed, at $f=0.5$, $N=64$ the whole energy
range is close to $E_N-E_1\approx 32$ being much larger
than the range $E_N-E_1\approx 2$   at $f=0$.
We argue that such a slow diffusion in energy
is at the origin of the approximate BE distribution
found in numerical simulations with the 
Bunimovich stadium [32] and the Sinai oscillator [33]
which have a very broad energy range
and the time of numerical simulations was not very
high due to the complexity of the integration of GPE.

\begin{figure}[H]
\begin{center}
\includegraphics[width=0.42\textwidth]{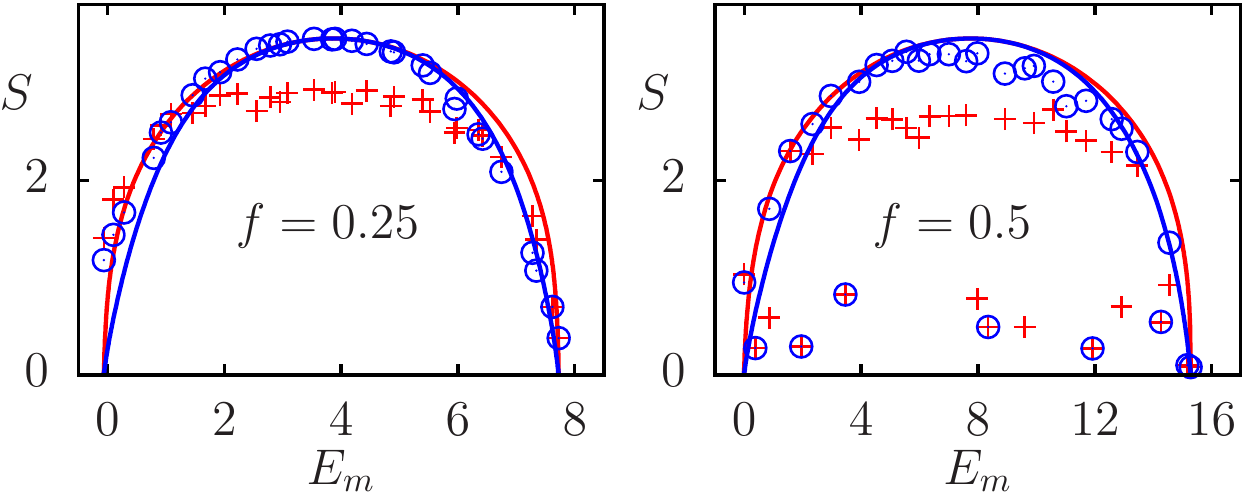}%
\end{center}
\caption{\label{figS13}
\label{fig_DIAG_SE}
As Fig.~1 for the case of an RMT plus extra diagonal matrix elements 
$fn$ with parameter $f=0.25$ or $f=0.5$ for $\beta=1$ and $N=32$. 
The data points correspond to the averaging time 
$2^{26}\le t\le 2^{27}$ (blue $\circ$ for $f=0.25,0.5$),
$2^{19}\le t\le 2^{20}$ (red $+$ for $f=0.5$) and 
$2^{14}\le t\le 2^{15}$ (red $+$ for $f=0.25$).
See also FIGURE NOTES of Fig.~\ref{figS13} and Fig.~\ref{figS14}.
}
\end{figure}

\begin{figure}[H]
\begin{center}
\includegraphics[width=0.42\textwidth]{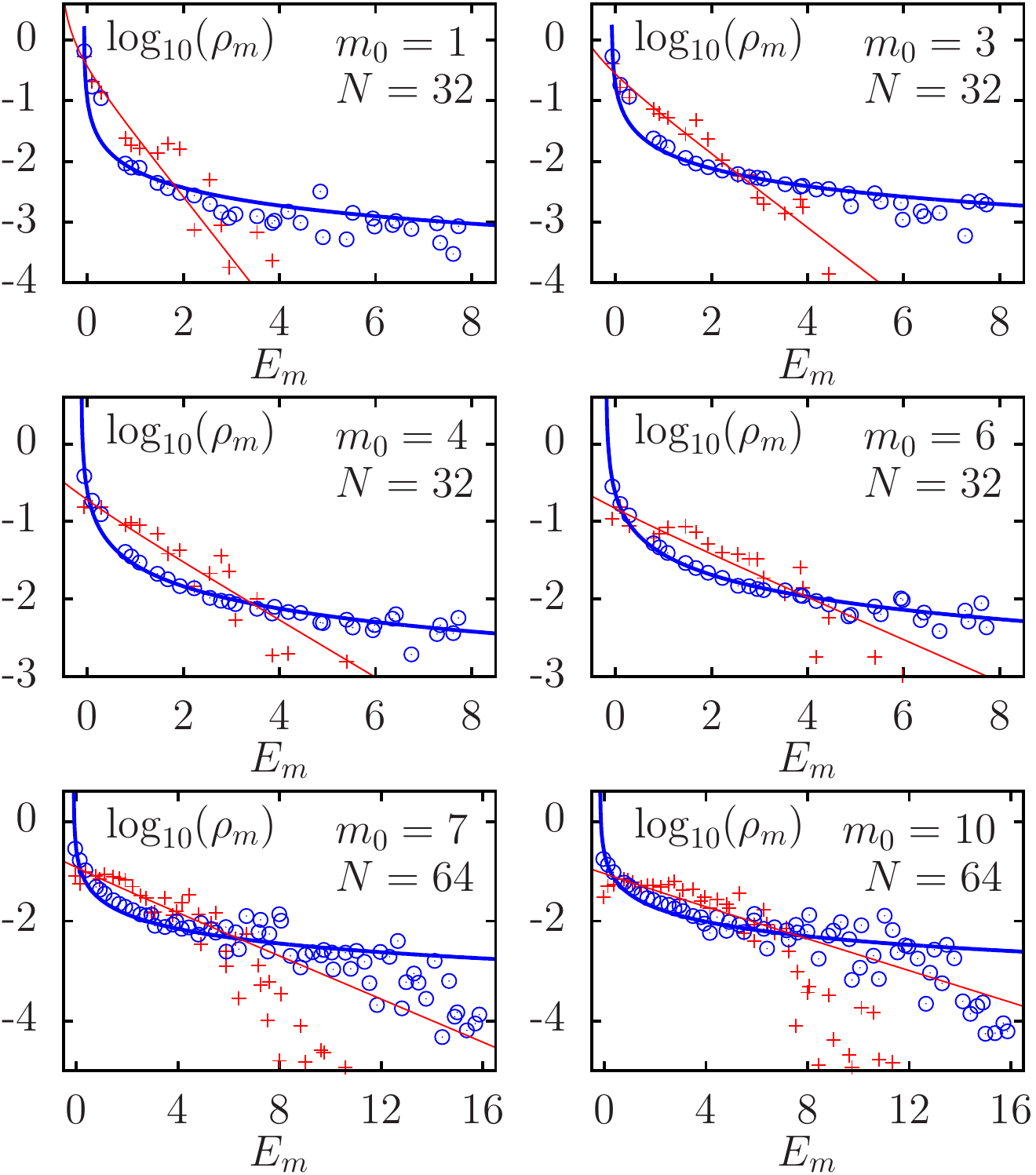}%
\end{center}
\caption{\label{figS14}
\label{fig_DIAG_states}
As Fig.~3 for the case of an RMT plus extra diagonal matrix elements 
$fn$ with parameter $f=0.25$ for $\beta=1$, $m_0=1,3,4,6$ ($N=32$) 
or $m_0=7,10$ ($N=64$).
The data points correspond to $\rho_m$ obtained by the averaging time 
$2^{26}\le t\le 2^{27}$ (blue $\circ$), 
$2^{19}\le t\le 2^{20}$ (red $+$; for $N=64$) and 
$2^{14}\le t\le 2^{15}$ (red $+$; for $N=32$). 
At longer times $t=2^{27}$ the states are (quite) well 
thermalized according to the EQ case (with somewhat stronger fluctuations 
for $N=64$). However, at the intermediate time 
scale the values of $\rho_m$ are closer to the BE line thus 
explaining that the corresponding entropy values are also closer to 
the BE curve.
See also FIGURE NOTES of Fig.~\ref{figS13} and Fig.~\ref{figS14}.
}
\end{figure}

{\bf Scaling of Lyapunov exponent and chaos border}

The numerical results presented in Figs. 5, \ref{figS6}-\ref{figS10}
are reasonably well described by the scaling relation:

\lformula{\label{eq_lyap} \lambda \sim \beta^{\eta}/N^\nu , \; \eta =3/2, \; \nu=2 .}

Indeed, the fits of data give values $\eta =1.52$ and $\nu =1.89$
being close to (\ref{eq_lyap}) and we assume that in the limit of large $N$ and small $\beta$
we will have the exponents of (\ref{eq_lyap}).

We find that most states with localized initial conditions 
($C_m(t=0)=\delta_{m,m_0}$) have zero Lyapunov exponents
at our smallest value $\beta =0.02$ (with a few exceptions
due to strong quasi degenerate levels as discussed above). But at the same time
the Lyapunov exponent is positive for random initial configurations
with random and uniform initial values of $C_m$ (which gives 
automatically an initial energy close to the energy band center).
All such states have approximately the same values of
$\lambda$ indicating that the measure of the 
chaotic component is close to unity.
At present, we cannot say what is the precise chaos border $\beta_c$
for such states. For the moment, we do not have theoretical arguments for
the found dependence (\ref{eq_lyap}).

We only note that equation (1) for the time evolution
can be rewritten in the basis of linear eigenmodes
(see eqs. Eq.(3) in [27] or Eq.(2) in [28]).
In this representation the transitions between modes
are induced only by $\beta$-terms with 4-mode interaction
(or 4-wave interaction)
$C^{\phantom{*}}_{m_1} C^{\phantom{*}}_{m_2} C^*_{m_3} C^*_{m}
\exp[-i(E_{m_1}+E_{m_2} - E_{m_3} - E_m)t]$.
In the RMT case the amplitudes of this interaction
have a typical value $Q \sim  1/N^{3/2}$ (see also [27,28]).
Thus the lowest energy difference between these 4 energies
is of the order of $\delta E \sim 1/N^2$ that can be
at the origin of $\nu \approx 2$ and rather low
chaos border with $\beta_c < 0.02$.
We note that the same estimate for $\delta E$
remains valid even in presence of the diagonal term $f n$
that stress the importance of 4-mode interactions.

However, the above estimates remain insufficient and the understanding
of the relation (\ref{eq_lyap}) requires further studies.

\section{Dynamical thermalization in multimode optical fibers}

Very recent remarkable experiments (published at 8 February 2023; 
after the submission date 22 December 2022 of our work)
with multimode optical fibers (MMF) [45]
demonstrated   dynamical thermalization in
MMF with negative temperature. It is 
stressed there that this is a dynamical thermalization 
resulting from pure Hamiltonian dynamics 
without an external thermal bath [45]. The equilibrium state
is a thermal state with energy equipartition over fiber modes
described by the EQ ansatz (2) also known in
optics as Rayleigh-Jeans distribution.
As we pointed out in Eq,(2) the EQ ansatz
is a limiting case of the BE ansatz when the temperature is large compared to
$E_m -\mu$ in the BE exponent. This can be considered
as the case when the field has many photons of linear modes.

There is a significant literature with discussions, numerical simulations
and experiments on dynamical thermalization in MMF
(see e.g. Refs.~S1,S2,S3,S4,[45]). The emergence of Rayleigh-Jeans distribution
is explained in the frame work of the weak turbulence approach
(see Refs.~S5,S6,S1,S2). However, it should be pointed out that
the weak turbulence theory (see Refs.~S5,S6) assumes
an existence of a certain weak randomizing force that disappears
in the final equilibrium state. This is in direct contradiction
with the dynamical Hamiltonian equations leading to the
equilibrium thermal state. In fact it is clear
that the origin of dynamical thermalization in MMF
is dynamical chaos and its exponential instability of
motion is related to a positive maximal Lyapunov exponent.
However, strangely enough no notion of dynamical chaos and
Lyapunov exponent appeared in theoretical arguments
of Refs.~S1,S2.S3.S4. Also from the theory of chaos
it is clear that no thermalization appears 
if the nonlinear perturbation is sufficiently weak
and below the chaos border (KAM integrability, see Refs.[5-8]).
In fact, we should note that in contrast to our RMT case the spectrum
of MMF discussed in Refs.~S1-S4,[45] has a form
$E \propto (m_x + m_y +const)$ thus with 
exact degenerate energy levels for the lowest 45 modes
considered practically in all MMF cases (and also in [45]).
As was shown in [26,40] for such a case with degeneracy of modes
the KAM theory is not valid and dynamical chaos
appears at an arbitrarily small nonlinear perturbation.
However, such chaos is localized only on degenerate modes
and does not lead to dynamical thermalization
over all modes.

Another interesting note about dynamical thermalization in MMF
experiments is about the validity of BE or EQ ansatz (2).
It is possible to assume that the light waves are classical
and then one should observe the EQ or  Rayleigh-Jeans distribution
over modes. However, the real life is
of course described by quantum mechanics with
second quantization of photons and their interactions
that should lead to the Bose-Einstein distribution (BE ansatz (2)).
It is possible that in MMF experiments the number of photons was very large,
dynamical temperature was high and the BE distribution 
was transferred to its classical limit with the EQ ansatz (or Rayleigh-Jeans).
However, it is interesting to know if MMF can operate
in a quantum regime with the BE thermal distribution.

We also point out that all discussed MMF systems
have very simple integrable spectrum with $E \propto (m_x + m_y +const)$
being rather far from the RMT spectrum which corresponds to a generic case.
Of course, it is difficult to realize such an RMT case with MMF.
However, it is possible to have cases when a fiber cross-section
have a form of a chaotic billiard. It may be the Bunimovich stadium 
(two semi-circles connected by two parallel straight lines),
or a circle with a line cut. In such systems the classical dynamics 
is chaotic and the level spacing statistics is the same as for RMT [24,25].
So we assume that such sections can be realized technologically
thus allowing to study nonlinear effects for MMF in a regime of quantum chaos.

\section{Generic features of dynamical thermalization in the NLIRM model}

The emergence of dynamical thermalization and its properties appeared 
as far as 150 years ago in 1872 in the work of Boltzmann who established 
the foundations 
of statistical mechanics and thermalization from dynamical equations [1]
(see also the related Boltzmann-Loschmidt dispute [2,3,4]).
The first attempt to obtain dynamical thermalization
in a nonlinear oscillator system, known as the FPU problem [10],
was not successful due to certain specific features of the FPU model.

In this work, we considered the NLIRM model (1),
which describes the classical dynamics of nonlinear oscillators, 
coupled by a Gaussian random matrix, and in which a moderate nonlinearity 
leads to the emergence of dynamical chaos followed by 
the classical dynamical thermal distribution Eq. (2) corresponding 
to the energy equipartition between oscillator modes of
the unperturbed linear system. 

We argue that, in contrast to the FPU problem [10],
our NLIRM model captures the generic features of
linear oscillator systems with moderate nonlinear
interactions between linear eigenmodes. 

First, the statistical classical theory given above in section I, 
is very generic and applies to generic linear couplings and 
generic interactions as long as we have the two integral 
of motions and as long as the system is sufficiently chaotic to 
ensure thermalization. 

Furthermore, without the nonlinearity the oscillators are described by 
Random Matrix Theory (RMT) which captures
the generic features of such diverse quantum systems as complex atoms, 
molecules and nuclei,
mesoscopic electronic systems and systems of quantum chaos [21,22,23,24,25].

In this work, we mostly used a nonlinear onsite interaction 
which is broadly used in condensed matter systems and is known
as the Hubbard interaction (see e.g. Ref. S7). We showed that
this interaction leads indeed to the EQ dynamical thermal distribution (2), 
perfectly confirming the theory of section I, 
not only for the linear oscillator system described by the RMT model 
but also by the DANSE model studied previously in [27,28,29] provided 
the iteration time is sufficiently long (see Figs.~S11, S12). 
This confirms the generic properties of the NLIRM model concerning 
{\em the linear oscillator couplings}. 

However, one can question if this model is also generic concerning 
the specific form of the onsite interaction and if the latter 
captures the generic features of dynamical thermalization in the NLIRM model.
To study this question, we have also considered 
two modified interaction models 
which are not restricted to onsite interactions only and which 
correspond (i) to nearest neighbors interactions (NNI) 
and (ii) to long range ``Coulomb type'' interactions (COULI). 
In these models the wavefunction evolution is described by the equation
\begin{align}
\label{eq1modif}
  i\hbar{\partial\psi_n(t) \over\partial t} =&  \sum_{n'=1}^N H_{n,n'} 
\psi_{n'}(t) \\
\nonumber
&+   \beta \left(\sum_j V_j \vert\psi_{n+j}(t)\vert^2\right) \psi_n(t)\ .
\end{align}
with interaction couplings $V_j=1$ for $j=-2,-1,0,1,2$ and $V_j=0$ 
for other $|j| >2$ in the NNI case
and $V_j = 1/(1+|j|)$ for $ -N/2 +1 < j <  N/2 +1$ in the COULI case;
the linear term with $H_{n,n'}$ (taken as a Gaussian random matrix) 
remains unchanged. In (\ref{eq1modif}), if $n+j<0$ or $n+j\ge N$ 
we apply periodic boundary conditions, i.e. $n+j\to n+j+N$ if $n+j<0$ 
and $n+j\to n+j-N$ if $n+j\ge N$. 
One can easily verify that for these types of interactions, we also have 
two integrals for motion being the conserved norm $1=\sum_n |\psi_n(t)|^2$ 
and the conserved classical energy which now reads~:
\begin{align}
\label{eq2modif}
E = &\sum_n \Big(<\psi_n(t)|\hat{H}|\psi_n(t)> \\ 
&+\frac{\beta}{2} \vert \psi_n(t) \vert^2
\sum_j V_j \vert \psi_{n+j}(t)|^2\Big)\ .
\nonumber
\end{align}
Therefore, the statistical classical theory given above in section I, equally 
applies to these kind of interactions. 

Furthermore, the considerations of section IIB can be generalized 
for these interactions. In particular in absence 
of the linear coupling (if $H_{nn'}=0$) the pure nonlinear dynamics 
conserves the individual values of $r_n=|\psi_n(t)|=$ const. and 
only the phases $\theta_n(t)$ evolve such that 
(for the pure nonlinear dynamics) we have:
\begin{align*}
\psi_n(t)=e^{-it\beta\sum_j|\psi_{n+j}(0)|^2}\,\psi_n(0)\ .
\end{align*}
This point is important to justify the use of the symplectic integrator 
which requires to compute the exact exponential $\exp(tB)$ 
of the operator $B$ corresponding to the nonlinear term 
(see section IIB).
We have verified that (i) the 4th order symplectic integrator, applied to both 
modified interaction models, still produces results such that the (global) 
error of the method scales with $(\Delta t)^4$ and (ii) that the classical 
energy (\ref{eq2modif}) is indeed conserved with small numerical variations 
$\sim 10^{-9}$-$10^{-8}$ for $\Delta t=0.1$.

\begin{figure}[H]
\begin{center}
  \includegraphics[width=0.42\textwidth]{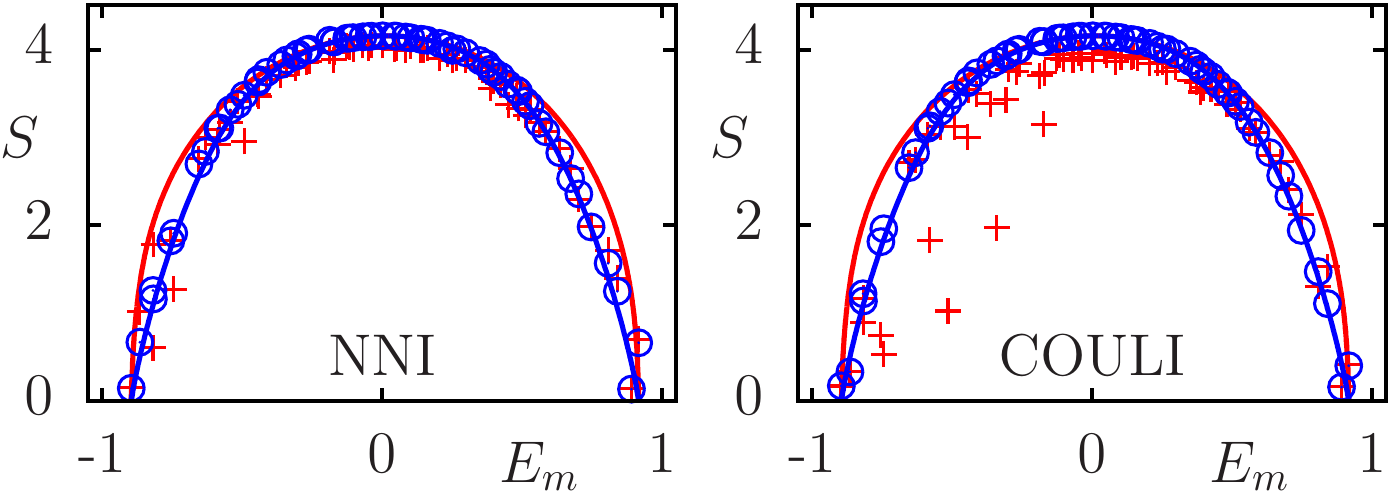}%
\end{center}
\caption{\label{figS15}
Dependence of entropy on energy $S(E)$ for both modified interaction 
models NNI, COULI, parameters $\beta=1,\,N=64$ and localized 
initial conditions (as in Fig.~1). 
The entropy S is computed from $\rho_m$ obtained by the time average 
$2^{23}\le t\le 2^{24}$ (blue $\circ$) or 
$2^{11}\le t\le 2^{12}$ (red $+$). Both panels 
have to be compared with Fig. 1(c) which corresponds to the same values 
of $\beta,N$ and average interval for $t$ but for the onsite interaction. }
\end{figure}

\begin{figure}[H]
\begin{center}
\includegraphics[width=0.42\textwidth]{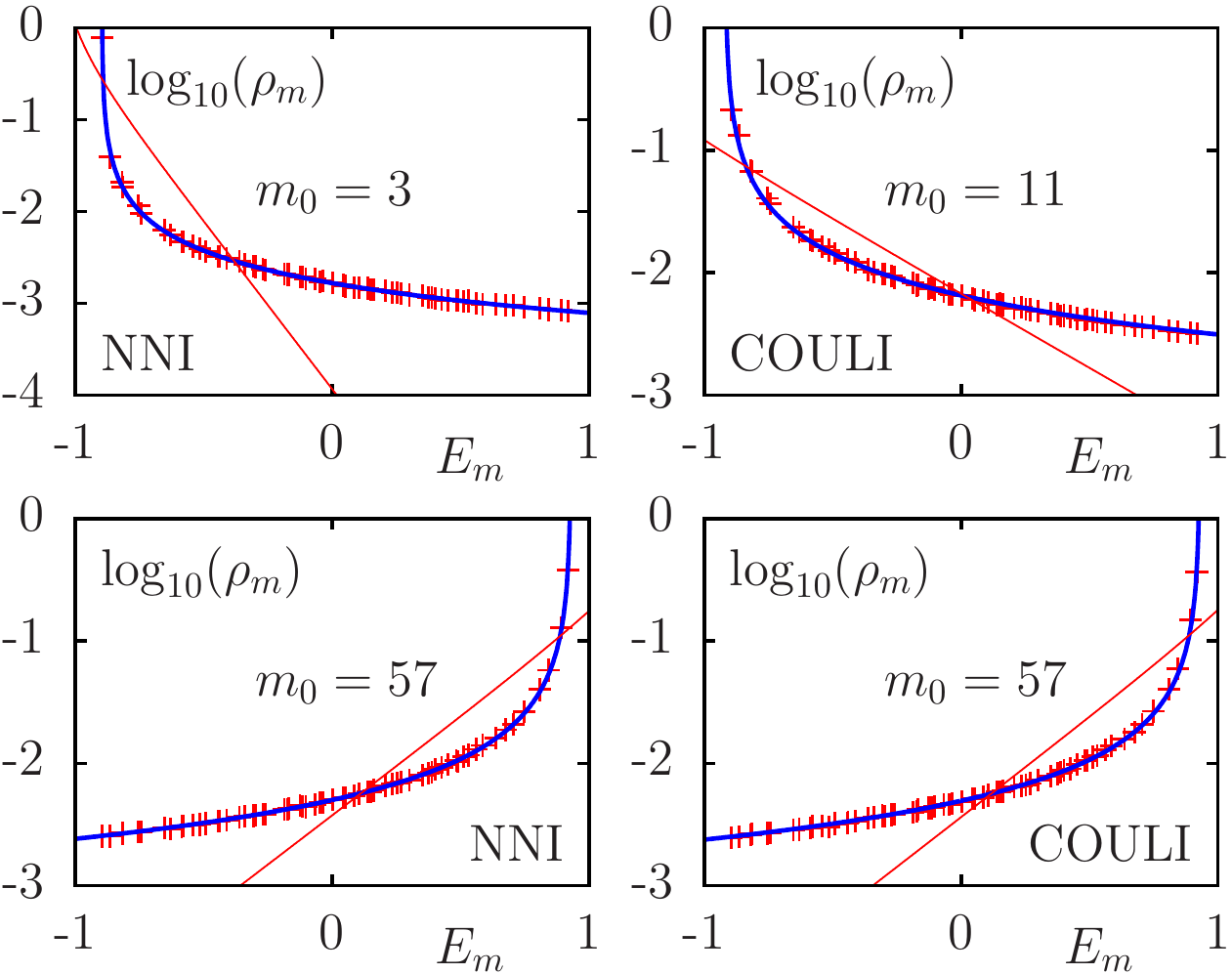}%
\end{center}
\caption{\label{figS16}
Dependence of $\rho_m$ on $E_m$ for both modified interaction models 
NNI and COULI and two states for each case 
with initial state $m_0=3,57$ for NNI and $m_0=11,57$ for COULI
obtained by an time average in the interval $2^{23}\le t\le 2^{24}$ 
(similar states and same parameters, $\beta=1,N=64$, as in 
Fig.~3 except for the modified interaction model). 
As in Fig.~3 the blue curve shows theory of the 
EQ ansatz with with $\rho_{\rm EQ}(E)=T/(E-\mu)$ and 
the red line shows the BE ansatz 
 $\rho_{\rm BE}(E)=1/(\exp[(E-\mu)/T]-1)$ with 
$\mu$ and $T$ determined from the norm and energy conservation 
as explained in the main text below Eq. (2). 
}
\end{figure}

The entropy dependence on energy $S(E)$ shown in Fig.~\ref{figS15} 
clearly confirms for both modified interaction models a thermalization 
to the classical EQ ansatz. The data points have to be compared with 
Fig.~1(c) which corresponds to the same values 
of $\beta,N$ and same time average intervals for $t$ 
but for the onsite interaction. The secondary set of data points 
for the reduced time interval, $2^{11}\le t\le 2^{12}$, is actually 
closer to the theoretical EQ-curve as compared to Fig.~1(c) showing that the 
thermalization time scale is even reduced, in particular for the NNI case.
For the longer time scale, $2^{23}\le t\le 2^{24}$, 
the data points lie nearly perfectly on the theoretical EQ-curve. 
Actually, our numerical data show that, for $\beta=1$, the thermalization 
is already very good for $t\ge 2^{15}$. 

Fig.~\ref{figS16} shows two examples for each modified interaction 
model NNI and COULI of the dependence of $\rho_m$ on $E_m$ for 
similar initial values $m_0$ as in Fig.~3. Also here the data matches 
perfectly the theoretical EQ-curves. More detailed figures 
for the full set of initial conditions, both modified interaction 
models, and $\beta=0.5,1,\,N=64$ are available at [39]. 

Therefore, the results presented in both figures clearly show that 
for both modified interaction models governed by Eq.~(\ref{eq1modif}), 
the  steady-state of the system is still very well described by the 
dynamical thermal distribution corresponding of Eq. (2) for the EQ case 
(with $T$ and $\mu$ determined by two implicit equations as explained 
below Eq. (2)). 

The physical reasons why a modification of the interaction range
does not affect the steady-state thermal distribution are 
(i) the theory of section I does not depend on the particular 
choice of the interaction, as long as it mixes the linear modes 
and (ii) the generic features of the linear RMT term corresponding 
to ``ergodic linear oscillator eigenmodes'' 
(i.e. ``ergodic'' in one particle quantum/oscillator space)
such that all types of moderate interactions lead to a nonlinear coupling of 
these modes with randomly fluctuating amplitudes (the same 
holds for the DANSE model if the ``linear quantum'' localization length 
is comparable to  the system size $N$, see also [27,29]).

We also point out that the dynamical thermal distribution
EQ (2) has been observed in experiments with multimode optical fibers
(see [45], Ref. S3, Ref. S4). 
\bigskip

------------SupMat References---------------------

\bigskip
Ref.S1. P.~Aschieri, J.~Garnier,  C.~Michel,  V.~Doya, and A.~Picozzi,
  {\it Condensation and thermalization of classsical optical waves in a waveguide},
  Phys. Rev. A {\bf 83}, 033838 (2011).

  Ref.S2. K.~Baudin , A.~Fusaro, K.~Krupa, J.~Garnier, S.~Rica, G.~Millot, and A.~Picozzi,
  {\it Classical Rayleigh-Jeans condensation of light waves: pbservation and
    thermodynamic characterization}, Phys. Rev. Lett. {\bf 125}, 244101 (2020).

  Ref.S3. E.V.~Podivilov, F.~Mangini,  O.S.~Sidelnikov,  M.~Ferraro,  M.~Gervaziev, 
  D.S.~Kharenko, M.~Zitelli, M.P.~Fedoruk, S.A.~Babin, and S.~Wabnitz,
  {\it Thermalization of orbital angular momentum beams in multimode optical fibers},
  Phys. Ref. Lett. {\bf 128}, 243901 (2022).

  Ref.S4. F.~Mangini, M.Gervaziev,  M.~Ferraro,  D.S.~Kharenko,
  M.~Zitelli,  Y.~Sun, V.~Couderc,  E.V.~Podivilov, S.A.~Babin, and S.Wabnitz,
  {\it Statistical mechanics of beam self-cleaning in GRIN multimode optical fibers},
  Optics Express {\bf 30(7)}, 10850 (2022).

  Ref.S5. V.E.~Zakharov, V.S.~L’vov, and G.~Falkovich, {\it Kolmogorov
  spectra of turbulence I}, Springer, Berlin, (1992).

  Ref.S6. S.~Nazarenko, {\it Wave turbulence}, Lectures Notes in Physics,
  Springer, New York (2011).
  
  Ref.S7. A.~Altland, and B.~Simons, {\it Condensed Matter Field Theory},
  p.58 Cambridge Univ. Press, Cambridge UK, (2006).


\end{document}